\documentclass[preprint, review, 12pt]{elsarticle}
\usepackage{graphicx}
\usepackage{amsmath}
\usepackage{textcomp}
\usepackage{amssymb}
\usepackage{relsize}
\journal{Journal of Industrial and Engineering Chemistry}

\begin{document}

\begin{frontmatter}
\title{Formation characteristics of Taylor bubbles in power-law liquids flowing through a microfluidic co-flow device}
\author{Somasekhara Goud Sontti}
\author{Arnab Atta\corref{cor1}}
\cortext[cor1]{Corresponding author. Tel.: +91 3222 283910}
\ead{arnab@che.iitkgp.ac.in}
\address{Multiscale Computational Fluid Dynamics (mCFD) Laboratory, Department of Chemical Engineering, Indian Institute of Technology Kharagpur, West Bengal 721302, India}

\begin{abstract}
	
Formation and dynamics of Taylor bubble in power-law liquids flowing through a circular co-flow microchannel are numerically investigated using coupled level set and volume-of-fluid method. Aqueous solutions of polyacrylamide (PAAm) are used as power-law liquids. Influences of PAAm concentration, gas-liquid velocities, and surface tension on bubble characteristics are explored. Various mechanism of bubble breakup are observed in different concentration of PAAm. Based on the bubble length with respect to the channel diameter, two different flow regimes are identified. Flow pattern maps are constructed based on inlet velocities, and scaling laws are proposed to estimate the bubble length.
\end{abstract}


\end{frontmatter}

\section{Introduction}

Taylor flow in microfluidics is categorized as one of the critical two-phase flow patterns, where capsular bubbles, mostly termed as Taylor bubbles, are formed in a continuous stream of liquid. The equivalent diameter of such a bubble is typically larger than the same of the associated microchannel. There exists a thin layer of liquid near the channel wall, which surrounds the Taylor bubble, and two consecutive bubbles are separated by a liquid slug \citep{taylor1961}. Due to reduced axial, and enhanced radial mixing \citep{gunther-2005, abiev2013bubbles}, analysis of Taylor flow is of paramount importance in the context of multiphase reactions \citep{dang2013reactivity,liedtke2016external} and micromixing \citep{zeng2012novel,xiao2016numerical}. Moreover, in several applications, the Taylor bubble size and flow behavior need to be controlled and precisely manipulated. Numerous studies have delineated various flow regimes \citep{tripl-1999,pohore-2008,zaloha-2012,park2009development}, and have proposed flow maps for wide-ranging microchannel configuration with gas\textendash liquid systems. Although, these attempts had limited applicability, but it is corroborated that Taylor flow regime occupies a significant part of any flow map. Over the years, considerable attention has been devoted to understand the bubble formation mechanism in various microchannel designs e.g., co\textendash flow \citep{lu-2015, wang2015speed}, T\textendash junction \citep{leclerc2010gas, peng2015,ma2017breakup}, and flow\textendash focusing \citep{lua-2016,dang2012preparation} devices. \citet{goel-2008}, and \citet{shaods-2008} numerically studied bubble formation in circular capillaries using volume\textendash of\textendash fluid (VOF) technique and investigated various influencing parameters, such as inlet conditions, superficial velocities, capillary diameter, and wall contact angle. \citet{chen-2009} employed level set (LS) method to study the Taylor bubble formation in a co\textendash flow configuration. \citet{fu-2015} reviewed numerous experimental and numerical studies on bubble formation, and breakup dynamics in microfluidic devices. Recently, \citet{Fletcher2016} proved conflicting results on liquid film thickness around a Taylor bubble, and provided several guidelines for reporting two\textendash phase flow systems in microchannels. Few researchers successfully captured the thin liquid film thickness in their numerical studies through near wall mesh refinement \citep{gupta-2009,jia-2016, sontti2017cfd}.

There is considerable amount of literature on Taylor bubble formation in Newtonian liquids. Interestingly, several fluids that are frequently encountered in industry and daily life including blood, protein, crude oil, polymer solution, etc. are likely to exhibit non-Newtonian behavior \citep{nghe2011microfluidics}. Moreover, in a two\textendash phase flow system, liquid properties play critical roles in heat and mass transfer studies \citep{shao2010mass, jia-2016}. Consequently, the bubble formation and breakup mechanism are complicated due to distinctive characteristics of non\textendash Newtonian fluids. Few experimental studies have depicted the effect of rheological properties on Taylor bubble formation in various microchannel configurations \citep{picchi-2015,labor-2015,mansour2015}. \citet{li2002} developed a theoretical model to understand the bubble formation in non\textendash Newtonian liquids (carboxymethylcellulose (CMC) and polyacrylamide (PAAm) solutions) by revising the model used for a Newtonian system. Predicted bubble volume and formation frequency were in reasonable agreement with the experiments. \citet{tang2012} studied flow characteristics of water and PAAm in untreated and hydrophobic microchannels. Their results showed that friction factors of PAAm solution were higher than theoretical values, and the hydrophobic microchannel was capable of reducing flow resistance as compared to untreated microchannels. Using a T\textendash junction microchannel, \citet{fu-2011} experimentally investigated bubble formation mechanism in non\textendash Newtonian liquids, and observed various flow patterns by adjusting the gas and liquid flow rates. Their results revealed that rheological parameters of non\textendash Newtonian fluid significantly influence bubble size, shape of the gaseous thread, and formation mechanism. \citet{wanga-2011} also demonstrated different flow patterns such as slug flow, slug\textendash annular flow, and annular flow in a T\textendash junction microchannel using a gas\textendash CMC system. 

With the help of micro-particle image velocimetry ($\mu$PIV), \citet{fu-2012b} analyzed the breakup of slender bubbles in non-Newtonian liquids flowing through a flow-focusing microchannel. Velocity and viscosity distributions around the gaseous thread were analyzed to understand the bubble breakup mechanism, and a scaling law was proposed to estimate the bubble size. \citet{mansour2015} studied two\textendash phase flow in a rectangular T-junction microchannel with different mass concentration of PAAm aqueous solutions. Their analysis also suggested significant effects of rheological properties on the flow pattern, bubble length, liquid slug length, bubble velocity, and frictional pressure drop. \citet{picchi-2015} studied the characteristics of air-CMC system in horizontal and inclined smooth pipes. The effect of pipe inclination and rheology of CMC were reported in terms of flow pattern maps obtained by visual observation. For various operating conditions, slug length, velocity, and frequency were also described. \citet{laborie2016} illustrated the effect of yield stress fluids on bubble formation in T\textendash junction, and flow\textendash focusing microchannels. They also provided a phase diagram for transient operation of bubble production in yield stress fluids. \citet{chen2013} developed a three-dimensional  numerical model for bubble formation in a T\textendash junction microchannel containing Newtonian and non\textendash Newtonian liquids using VOF method. Initially, their numerical model was verified for Newtonian liquids with in-house experimental visualization, and thereafter the model was extended for the power-law and Bingham fluids. However, there are limitations of the VOF method regarding surface tension force modeling, which limit its applicability in estimating smoothed physical properties across the interface. Improper formulation of balance between surface tension induced capillary force and pressure jump across the interface leads to the development of unphysical velocity near the interface, which is typically referred to as spurious currents \citep{harvie2006,popinet1999}. Several researchers have attempted to reduce spurious currents with different approaches \citep{aulisa2006,francois2006,popinet2009,guo2015}. \citet{sussman2000} developed a coupled LS and VOF (CLSVOF) technique by combining the advantages of both the methods. This strategy helps to utilize the advection of the VOF function for conserving the mass, and to smoothly capture the interface by calculating the radius of curvature from the LS function, simultaneously. 

Numerous studies have been reported using CLSVOF method in different applications, e.g. bubble rise in viscous liquids \citep{keshavarzi2014}, bubble formation on submerged orifices \citep{buwa2007}, droplet impact on a liquid pool \citep{ray2015}, influence of the fluid properties on Taylor bubble formation \citep{dang-2015}, axisymmetric droplet formation \citep{chakraborty2016}, droplet coalescence \citep{mino2016}, and demulsification \citep{kagawa2014}. Nevertheless, these studies are focused on the bubble/droplet behavior only in Newtonian liquids. Few CLSVOF studies have been reported on bubble generation in non\textendash Newtonian liquids \citep{fan2014,fan2016}, which suggest that this combined approach can accurately capture the sharp interface in non-Newtonian systems, as well. Although, such results show interesting perspectives on bubble breakup mechanism in non\textendash Newtonian fluids, it can be understood from those studies that further investigation is essential for analyzing Taylor bubble behavior in non\textendash Newtonian liquid flowing through a microchannel. In this article, we present a CLSVOF model to understand the gas\textendash non-Newtonian liquid flow in a circular co\textendash flow microchannel. We investigate the underlying physics of bubble formation and attempt to form flow regime maps by analyzing the bubble shape, velocity, and surrounding liquid film thickness in power-law liquids.    

\section{Coupled LS and VOF (CLSVOF) method}

Despite being mass conservative, the VOF method often results in spurious currents. This issue is identified as a consequence of an unbalanced representation of surface tension force and pressure variation across the interface. In contrast, LS method enables tracking of a smoother interface, however, it suffers from the mass conservation issue \citep{suss-1994}. In LS method, a signed distance function is used to identify the phases. In the present study, a coupled LS and VOF (CLSVOF) method is implemented to overcome the deficiencies of both LS and VOF methods.

\subsection{VOF method}

In VOF approach, a single set of conservation equations is solved for immiscible fluids \citep{hirt-1981}. The governing equations of the VOF formulation for multiphase flows are as follows:

\textbf{Equation of continuity:} 
\begin{equation}
\label{eq:mass_eqn}
\frac{\partial  \rho }{\partial t}  +  \nabla .  ( \rho  \vec{ U } ) =0
\end{equation}

\textbf{Equation of motion:}
\begin{equation}
\label{eq:mom_eqn}
\frac{ \partial (\rho \vec{ U })}{ \partial t} + \nabla.( \rho \vec{ U } \vec{ U }) = - \nabla P + \nabla.\overline{\overline \tau} + \vec{ F}_{SF}
\end{equation}

\begin{equation}
\label{eq:tau}
\overline{\overline \tau}= \eta \dot{ \gamma } = \eta (\nabla \vec {U} + \nabla { \vec {U} } ^{T})
\end{equation}

where $\vec{U }$, $\rho $, $\eta $, $P$, and $\vec{ F}_{SF}$ are velocity, density, dynamic viscosity of fluid, pressure, and surface tension force, respectively.

\textbf{Equation of VOF function:} 
The interface between two phases can be traced by solving the following continuity equation of the volume fraction in absence of any mass transfer between phases.  
\begin{equation}
\label{eq:vof_eqn} 
\frac{\partial \alpha _{q}}{\partial t} + (\vec{ U_q }.\nabla) \alpha _{q} =0
\end{equation}

where $\alpha_{q}$ is the volume fraction of \textit{q}th phase (gas or liquid). For a two-phase system, if the phases are represented by the subscripts (1 and 2), and the volume fraction of the phase 2 is known, then the density and viscosity in each cell are given by: 
\begin{equation}
\rho =  \alpha _{2}  \rho _{2}  + (1 - \alpha _{2})\rho _{1} 
\end{equation}
\begin{equation}
\eta =  \alpha _{2}   \eta  _{2}  + (1 - \alpha _{2}) \eta  _{1} 
\end{equation}

The volume fraction for the primary phase can be obtained from the following equation:
\begin{equation}
\sum  \alpha _{q} =1
\end{equation} 

\subsubsection{Surface tension force}

In VOF method, continuum surface force (CSF) model \citep{brack-1992} is typically used to define the volumetric surface tension force ($F_{SF})$ term in Eq. \ref{eq:mom_eqn}, as follows:
\begin{equation}
\label{eq:source_eqn}
\vec{F} _{SF} = \sigma  \begin{bmatrix} \frac{\mathlarger{\rho } \mathlarger{\kappa_{N}} \mathlarger{\nabla \alpha_1}}{ \mathlarger{\frac{1}{2 }} (\mathlarger{\rho_{1}+  \rho_{2}) }}  \end{bmatrix}    
\end{equation}

where $\kappa_{N}$ and $\sigma$ are the radius of curvature and the coefficient of surface tension, respectively. The interface curvature ($\kappa_{N}$) is calculated in terms of unit normal, $\hat{N}$, as:
\begin{equation}
\label{eq:normal_eqn}
\kappa_{N} = - \nabla  .  \hat{N}= \frac{1}{|\vec{ N}|}  \begin{bmatrix} \big( \frac{\vec{ N}}{ |\vec{ N}|} . \nabla\big) |\vec{ N}| - \big( \nabla  . \vec{ N}\big) \end{bmatrix}; \text {and} ~\hat{N} = \frac{\vec{ N}}{ | \vec{ N} | }
\end{equation}
where $\vec{N}$ is expressed as the gradient of phase volume fraction at the interface (Eq.~\ref{eq:vofnormal}): 
\begin{equation}
\centering
\label{eq:vofnormal}
\vec{N}= \nabla  \alpha _{q} 	
\end{equation}
Wall adhesion effect is also incorporated by defining a three-phase contact angle at the channel wall ($\theta_{W}$). Accordingly, the surface normal at the reference cell next to the wall is given by:
\begin{equation}
\label{eq:normal1_eqn}
\hat{N}=  \hat{N}_{W}  cos \theta_{W}  +  \hat{M}_{W} sin \theta _{W} 
\end{equation}

where $\hat{N}_W$ and $\hat{M}_W$ are the unit vectors normal and tangential to the wall, respectively \cite{fluent}. In VOF, it is difficult to capture the geometric properties (interface normal and curvature) from the VOF function whose spatial derivatives are not continuous near the interface. Such inaccurate calculations of geometric properties may lead to spurious currents. Therefore, the volumetric surface tension force ($F_{SF})$ term is modified with a continuous LS function to reduce spurious currents that helps in improving radius of curvature estimation, as mentioned in the subsequent section.   

\subsection{Equation of LS function:} 
\begin{equation}
\frac{\partial   \varphi }{\partial t}  +  	\vec{ U_q } .  \nabla  \alpha _{q}=0  
\end{equation}
where $\varphi $ and $\alpha$ are the LS function and volume fraction in the \textit{q}th cell, respectively. $\varphi$ is a function of position vector ($\vec{\chi}$) and time ($t$), which acts as the signed distance from the interface, as follows \citep{suss-1994}:   
\begin{equation}
	\varphi ( \vec{\chi },{{t} } ) = \begin{cases}d  & \text{if} \hspace{5px} \text{$\chi$ is  in the liquid phase}\\0 & \text{if} \hspace{5px} \text{$\chi$ is  in the interface}\\-d & \text{if} \hspace{5px} \text{$\chi$ is in the gas phase} \end{cases}  
\end{equation}
where $\textit{d}=d(\vec{\chi})$ is the shortest distance of a point $\vec{\chi}$ from interface at time $\textit{t}$. The fluid phase is identified based on the sign of the LS function. It takes positive values in the liquid region, and assigns negative values in the gas phase, whereas zero value is specified at the interface. To solve the governing equations (Eq. \ref{eq:mass_eqn} and Eq. \ref{eq:mom_eqn}), smoothed distribution of the fluid properties (density and viscosity) especially across the interface is required. In this work, both the fluids are assumed to be incompressible. This leads to adopting two different values depending on the sign of LS function in the solution domain. Therefore, the mixture physical properties are also calculated using a smoothed Heaviside function $\textit{H}(\varphi)$ across the interface, as follows \citep{sussman2000}:
\begin{equation}
\rho (\varphi )=\textit{H}(\varphi)\rho_{2}+(1-\textit{H}(\varphi))\rho_{1}  
\end{equation} 
\begin{equation}
\eta(\varphi)=\textit{H}(\varphi)\eta_{2}+(1-\textit{H}(\varphi))\eta_{1}
\end{equation}

The smoothed Heaviside function ($\textit{H}(\varphi)$) is defined as:
\begin{equation}
\textit{H}(\varphi)=\begin{cases} 0& \textit{ if} \hspace{5px} \varphi  < - a  \\  \frac{1}{2}[1+ \frac{ \varphi }{a} + \frac{1}{ \pi }sin( \frac{ \pi  \varphi }{a} ) ]  & \textit{ if} \hspace{5px} |  \varphi  |  \leq a\\ 
1 & \textit{ if} \hspace{5px} \varphi  > a
\end{cases}  
\end{equation}
where \textit{a} is the interface thickness.

Finally, the volumetric surface tension ($\vec{ F}_{SF} $) force (in Eq. \ref{eq:mom_eqn}) based on CSF method in CLSVOF is calculated as \citep{sussman1999}:
\begin{equation}
\vec{ F}_{SF} = \sigma  \kappa ( \varphi ) \delta ( \varphi ) \nabla \varphi 
\end{equation}

where $\kappa ( \varphi )$ and $\delta ( \varphi )$  are the interface curvature and the smoothed Dirac delta function, respectively, defined as:
\begin{equation}
\kappa ( \varphi )= \nabla .  \frac{ \nabla  \varphi }{ |  \nabla  \varphi  | } 
\end{equation}
\begin{equation}
\delta ( \varphi )=\begin{cases}0 & \textit{ if} \hspace{5px} |  \varphi  |  \geq a\\ \frac{1}{2a} (1+cos( \frac{ \pi  \varphi }{a} )) & \textit{ if} \hspace{5px} |  \varphi  |  < a\end{cases} 
\end{equation}


\subsubsection{Constitutive equation of continuous phase}

To investigate the bubble formation in non\textendash Newtonian liquid, a power-law model is considered for calculating the effective viscosity ($\eta_{eff}$) that is expressed as a function of shear rate \citep{chhabra2011}. For non\textendash Newtonian liquids, the shear stress can be written in terms of a non\textendash Newtonian viscosity, as follows:
\begin{equation}
\centering
\label{eq:nnvis}
\overline{\overline \tau} = \eta_{eff}(\dot{\gamma})\dot{\gamma}
\end{equation}
where  $\eta_{eff}$ is a function of all three invariants of the rate\textendash of\textendash deformation tensor. However, in power\textendash law model, the non\textendash Newtonian liquid viscosity ($\eta_{eff}$) is considered to be a function of only shear rate ($\dot{ \gamma }$) (Eq. \ref{eq:nnvis_eqn}).
\begin{equation}
\label{eq:nnvis_eqn}
\eta_{eff}(\dot{\gamma}) =K \dot{\gamma }^{n-1} 
\end{equation} 
where $K$ and $n$ are the consistency and power-law indices, respectively. The local shear rate ($\dot{ \gamma}$) is expressed as \citep{fluent}:
\begin{equation}
\label{eq:shearrate}
\dot{ \gamma } = \sqrt{ \frac{1}{2} (\nabla \vec {U} + \nabla { \vec {U} } ^{T})_{ij} (\nabla \vec {U} + \nabla { \vec {U} } ^{T})_{ji}} 
\end{equation} 

\subsection{Model implementation}
\begin{figure}[!ht]
	\centering
	\includegraphics[width=0.8\linewidth]{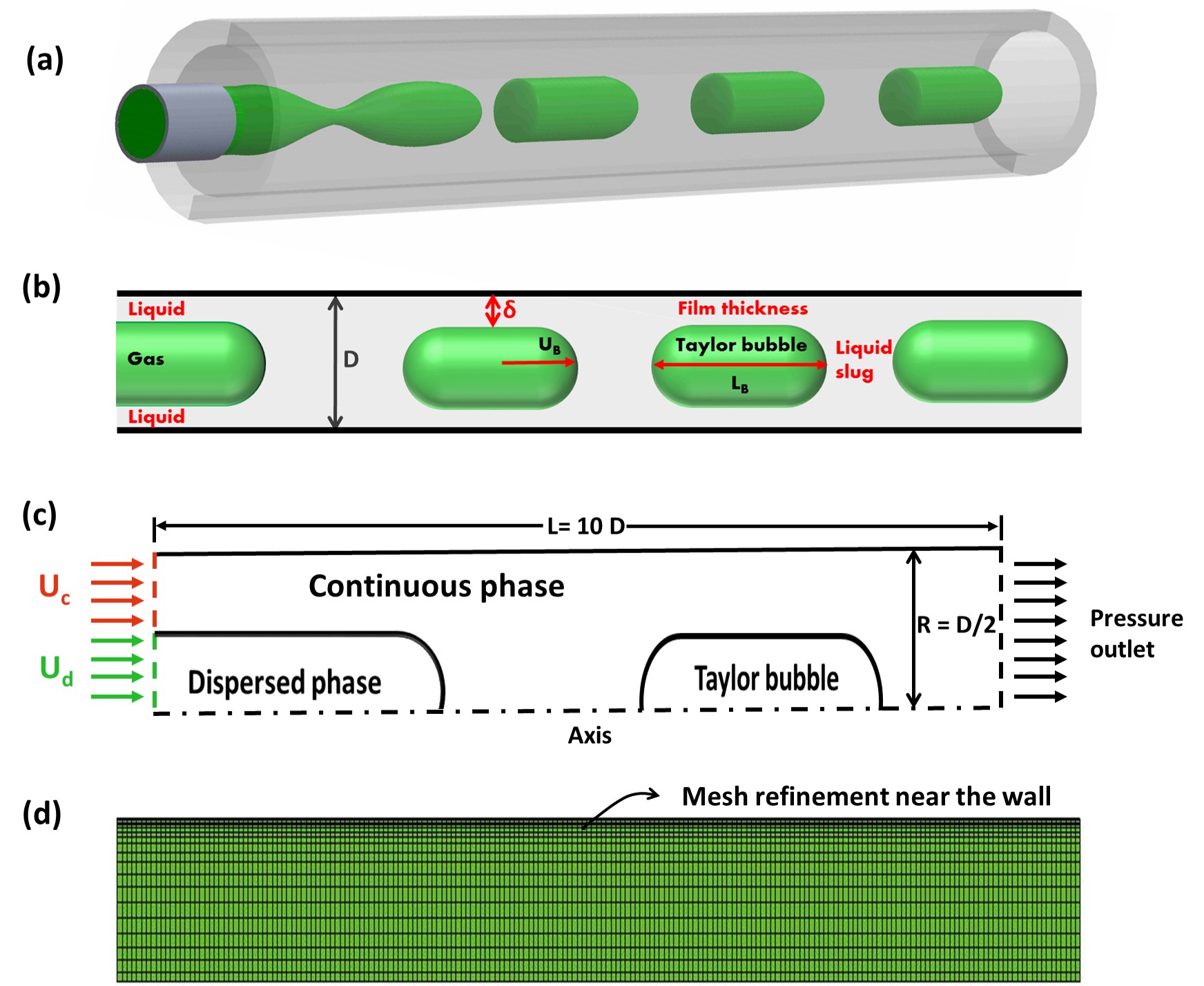}
	\caption{\label{fig:refine_mesh}Schematic of (a) 3D cross-sectional view of Taylor bubble formation in a circular co\textendash flow geometry, (b) 2D depiction of bubbles surrounded by a thin liquid film, and separated by liquid slug, (c) 2D axisymmetric representation of computational domain with imposed boundary conditions, and (d) mesh refinement near the wall.}
\end{figure}

Taylor bubble formation process in a circular co-flow microchannel is schematically depicted in Fig.~\ref{fig:refine_mesh}a.
In this study, a circular microchannel of diameter (D) 0.5~mm, and length of 10~D is considered. Gas phase inlet capillary diameter is taken as 0.35 mm assuming negligible wall thickness. Fig.~\ref{fig:refine_mesh}b shows two-dimensional planar view of the Taylor bubbles inside the microchannel including parameters such as, bubble length ($L_B$), bubble velocity ($U_B$), and surrounding liquid film thickness ($\delta$), which are analyzed to understand the air\textendash PAAm flow behavior. The computational domain is considered as a two-dimensional axisymmetric geometry (Fig.~\ref{fig:refine_mesh}c). Finite volume method based solver is used to solve aforementioned partial differential equations. Pressure implicit with splitting operators (PISO) algorithm is used to resolve the pressure\textendash velocity coupling in momentum equation \citep{issa1986}. The spatial derivatives of momentum and LS equations are discretized using the second-order upwind scheme \citep{barth1989}. The volume fraction is solved using piecewise linear interface construction (PLIC) algorithm \citep{holt2012}. Subsequently, variable time step and fixed Courant number (Co = 0.25) are considered for solving the governing equations. At the inlet, constant velocity boundary condition is imposed for both continuous (liquid) and dispersed (gas) phases. Recently, using $\mu$PIV, \citet{fu2016newtonian} investigated the velocity field distribution in various concentration of PAAm (0.10\% \textendash 1.25\%) flowing inside microchannels of different dimensions ranging from 400 $\mu$m to 800 $\mu$m. They clearly stated that the slippery phenomenon was not observed in their experiments as the sizes of microchannels were much greater than 10 $\mu$m. Moreover, \citet{vayssade2014} demonstrated that non-Newtonian flows in confined systems were dominated by slip heterogeneities only below a certain length scale. Interestingly, in the present study, similar channel dimension and PAAm solutions are used as considered by \citet{fu2016newtonian} in their experimental work. Therefore, the non-slip condition at the impermeable and confined wall of the microchannel is implemented, which is also in line with previous literature \cite{madadelahi2018,ren2015,cho2012,sang2009}. Pressure boundary condition is specified at the channel outlet by assigning zero gauge pressure. Quadrilateral mesh elements are utilized, and the grid independence study is initially performed using different mesh element sizes ranging from 3 $\mu m$ to 7 $\mu m$. In cases of coarser grids (e.g., 6 $\mu m$), diffusive interface is observed, and small satellite bubble is noticed in the liquid slug, as shown in Fig.~\ref{fig:V123}a. However, the results did not show any appreciable difference bubble length estimation under identical operating condition, as depicted in Fig.~\ref{fig:V123}b. Therefore, an element size of 5 $\mu m$ in the core of the channel is used for rest of the study. It is noteworthy that in the present study, liquid film thickness is distinctly captured in all cases with mesh refinement near the solid wall, as illustrated in Fig.~\ref{fig:refine_mesh}d. 

\section{Model validation}
\begin{figure}[!ht]
 	\centering
 	\includegraphics[width=0.8\linewidth]{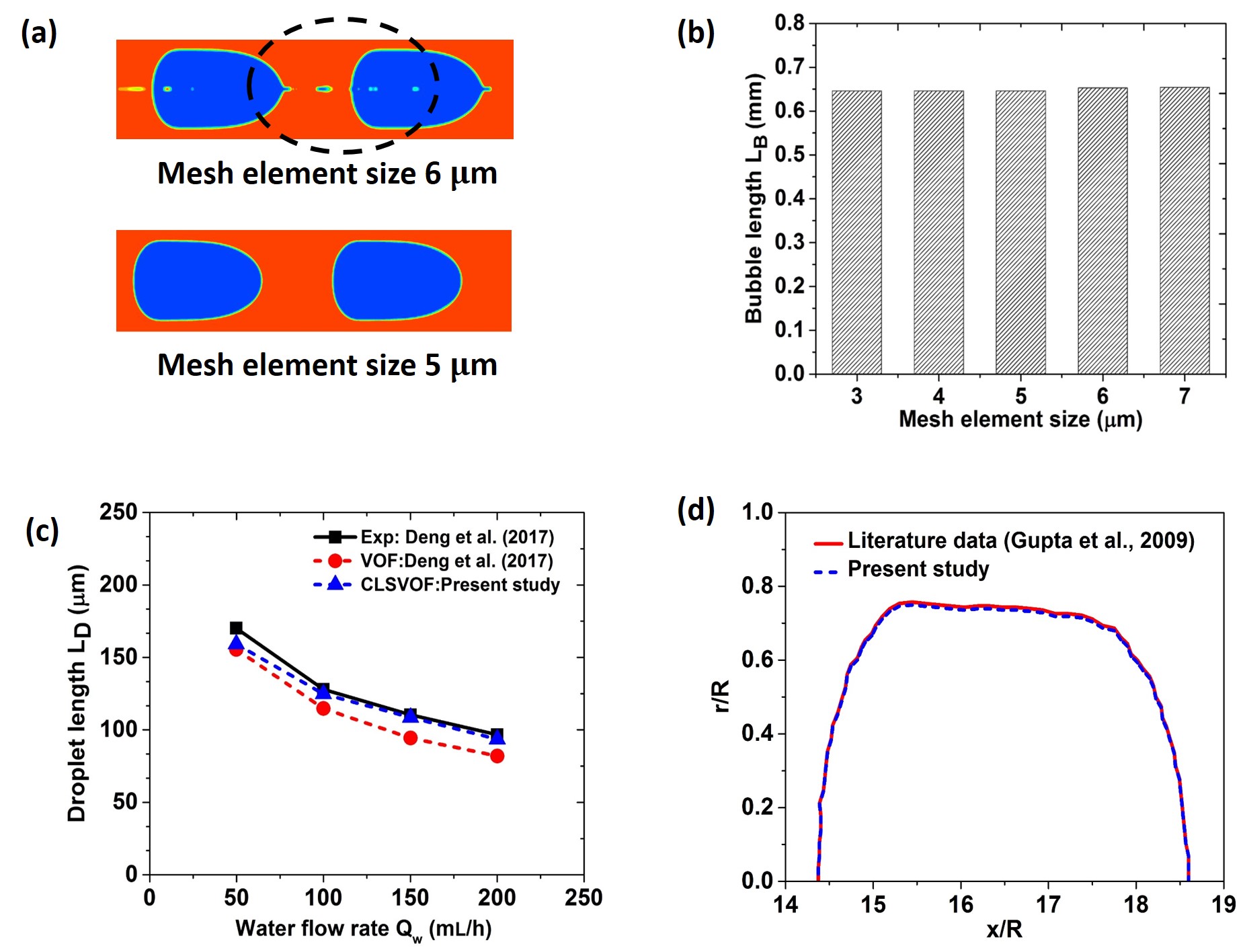}
 	\caption{\label{fig:V123} (a) Comparison of interface tracking by two different mesh element sizes, (b) grid independence study of the bubble length for air-PAAm~1.25\% system at $U_G$ = 0.5 m/s, $U_L$ = 0.5 m/s, (c) comparison of model predications with experimental and numerical (VOF) results of \citet{deng2017} at $Q_o$ = 0.03 $mL/h$, $\sigma$ = 19.45 mN/m, $\eta_{oil}$ = 49.50~mPa.s, and $\eta_{water}$= 1.04~mPa.s, and (d) comparison of Taylor bubble shape for air-water system at $U_G$ = 0.5 m/s, $U_L$ = 0.5 m/s with the results of \citet{gupta-2009}.}
 \end{figure}

Firstly, Newtonian two-phase (oil\textendash water) flow in a circular microchannel is studied to examine the validity of our developed CFD model. The results of CLSVOF method are compared with experimental and VOF simulation results of \citet{deng2017} under the same operating condition. Fig.~\ref{fig:V123}c shows the comparison of droplet length in a circular microchannel.
 The simulated droplet lengths are found to be in close agreement with the experimental data. These results affirm the correctness of our numerical model in predicting the behavior of droplet length in co\textendash flow device. A noticeable difference in droplet length is observed at a lower flow rate with a maximum deviation of 6\% from the experimental results, as shown in Fig.~\ref{fig:V123}c. Moreover, our simulation results appear to be better than that of the VOF method. This validation also establishes the efficacy of the developed model to forecast the bubble length better than the VOF simulations, as reported by \citet{deng2017}. The developed model is further verified for a gas-Newtonian liquid system by comparing the bubble shape with the results of \citet{gupta-2009}. It is evident from Fig.~\ref{fig:V123}d that the developed CLSVOF model accurately predicted the Taylor bubble shape, as well.

\section{Results and discussions}
In this work, different concentrations of aqueous PAAm solutions are considered as liquid phase, and the rheological properties of PAAm solutions are adopted from the experimental work of \citet{fu-2012b}. It was shown that in the range of 0.1\% to 1.25\% concentrations, PAAm aqueous solutions do not exhibit any elastic behavior \cite{fu-2012b}. Moreover, it has been reported that the considered range of PAAm concentration in this study features shear thinning power-law behavior \cite{dietrich2008passage, fu2016breakup, fu2011gas}. Values of consistency index ($K$), power\textendash index ($n$) and other physical properties of the liquid phase are listed in Table~\ref{tab:power_law_data_picchi}. It is interesting to note from Table~\ref{tab:power_law_data_picchi} that the surface tension and densities of all solutions are close to that of water because the concentrations of PAAm in the aqueous solutions are relatively low. Influences of various parameters namely, PAAm concentration, fluid inlet velocities, and surface tension ($\sigma$) are elaborated in the subsequent sections.

\begin{table}[h]
	\centering
	\small
	\caption{Rheological properties of PAAm aqueous solutions \citep{fu-2012b} used in this study.}
	\label{tab:power_law_data_picchi}
	\begin{tabular}{@{}lcccc@{}}
		\hline
		\begin{tabular}[c]{@{}l@{}}Liquid phase\\ (PAAm\textendash wt\% Conc.)\end{tabular} & \begin{tabular}[c]{@{}c@{}}Density, \\ $\rho$ ($kg/m^{3}$)\end{tabular} & \begin{tabular}[c]{@{}c@{}}Power\textendash law\\ index, n  (\textendash)\end{tabular} & \begin{tabular}[c]{@{}c@{}}Consistency \\ index, $K$ ($Pa.s^{n}$)\end{tabular} & \begin{tabular}[c]{@{}c@{}}Surface tension,\\  $\sigma$ (mN/m) \end{tabular} \\ \hline
		PAAm\textendash0.1\%   & 1000 & 0.49 & 0.34  &71.1 \\
		PAAm\textendash0.25\%   & 1000 & 0.41& 1.05 & 70.3 \\
		PAAm\textendash0.5\%  & 1000 & 0.36 & 2.87  & 69.6 \\
		PAAm\textendash0.75\%  & 1000 & 0.34 & 4.32& 69.3 \\
		PAAm\textendash1.0\%  & 1000 & 0.29 & 7.91 & 67.7 \\
		PAAm\textendash1.25\%  & 1000 & 0.26 & 10.85 & 67.2 \\
		\hline
		
	\end{tabular}
\end{table}

\subsection{Effect of PAAm concentration}
In this section, the effect of PAAm concentration on Taylor bubble formation, bubble length, its velocity, and liquid film thickness has been examined. To understand the viscous effect of power-law liquids on the two\textendash phase flow, effective viscosity \citep{chhabra2011} is calculated, and the influence of PAAm concentration on bubble generation is explained in terms of this effective viscosity of the continuous phase and its velocity distribution in the microchannel. Effective viscosities ($\eta_{eff}$) of these power-law liquids are calculated using Eq.~\ref{eq:effective_eqn}, as follows:    
%
\begin{equation}
	\label{eq:effective_eqn}
	\eta_{eff}=K\left(\frac{3n+1}{4n}\right)^{n}\left(\frac{8U_{L}}{D}\right)^{n-1}
\end{equation}
where $K$, $U_L$, $D$, and $n$ are consistency index, liquid velocity, diameter of the channel, and power\textendash law index, respectively.

\begin{figure}[h]
	\centering
	\includegraphics[width=\linewidth]{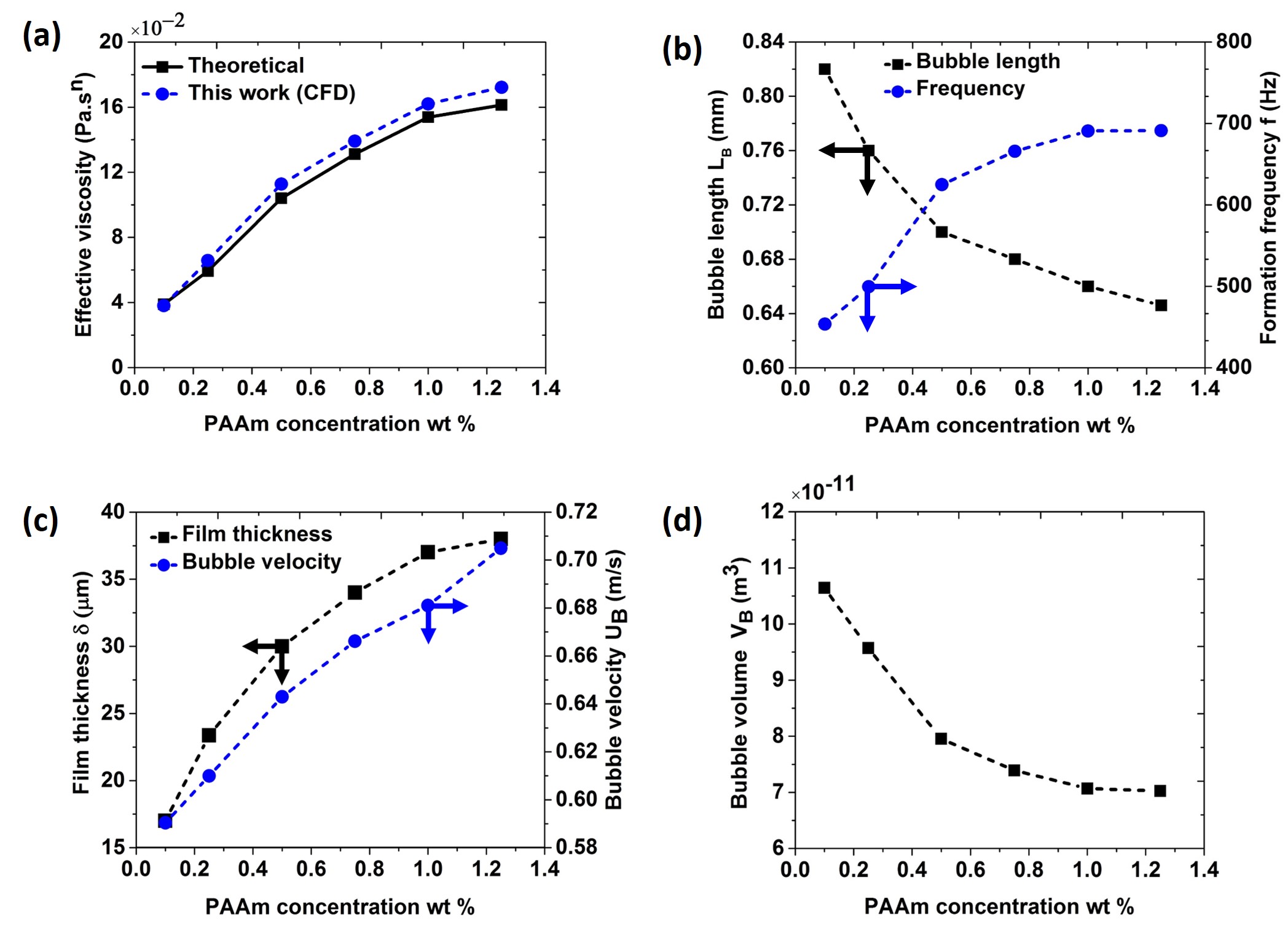}
	\caption{\label{fig:paam_11} Effect of PAAm concentration on (a) effective viscosity, (b) bubble length and formation frequency, (c) liquid film thickness and bubble velocity, and (d) bubble volume at $U_{L}$ = 0.5 m/s and $U_{G}$ = 0.5 m/s.}
\end{figure}

It is apparent from Eq.~\ref{eq:effective_eqn} that with increasing PAAm concentration, effective viscosity of the solution increases, as also shown in Fig.~\ref{fig:paam_11}a. Due to enhanced effective viscosity of the solution bubble length decreases, and the formation frequency increases with increasing PAAm concentration, which are depicted in Fig.~\ref{fig:paam_11}b. Similar observation was also experimentally reported by \citet{fu-2011}, who studied bubble formation in a T-junction microchannel with different concentrations of carboxylmethyl cellulose (CMC). In all cases, a thin film of the continuous phase around the Taylor bubble is precisely captured and the liquid film thickness ($\delta$) is analyzed. Fig.~\ref{fig:paam_11}c shows that film thickness around the bubble increases with increasing PAAm concentration, which consequently leads to increased bubble velocity \cite{breth-1961}. This is attributed to the enhanced viscous force at higher PAAm concentrations. Fig.~\ref{fig:paam_11}d shows the decrease in bubble volume with increasing PAAm concentration due to the combined reduction of bubble length and the width, as discussed earlier.

\begin{figure}[!ht]
	\centering
	\includegraphics[width=0.85\linewidth]{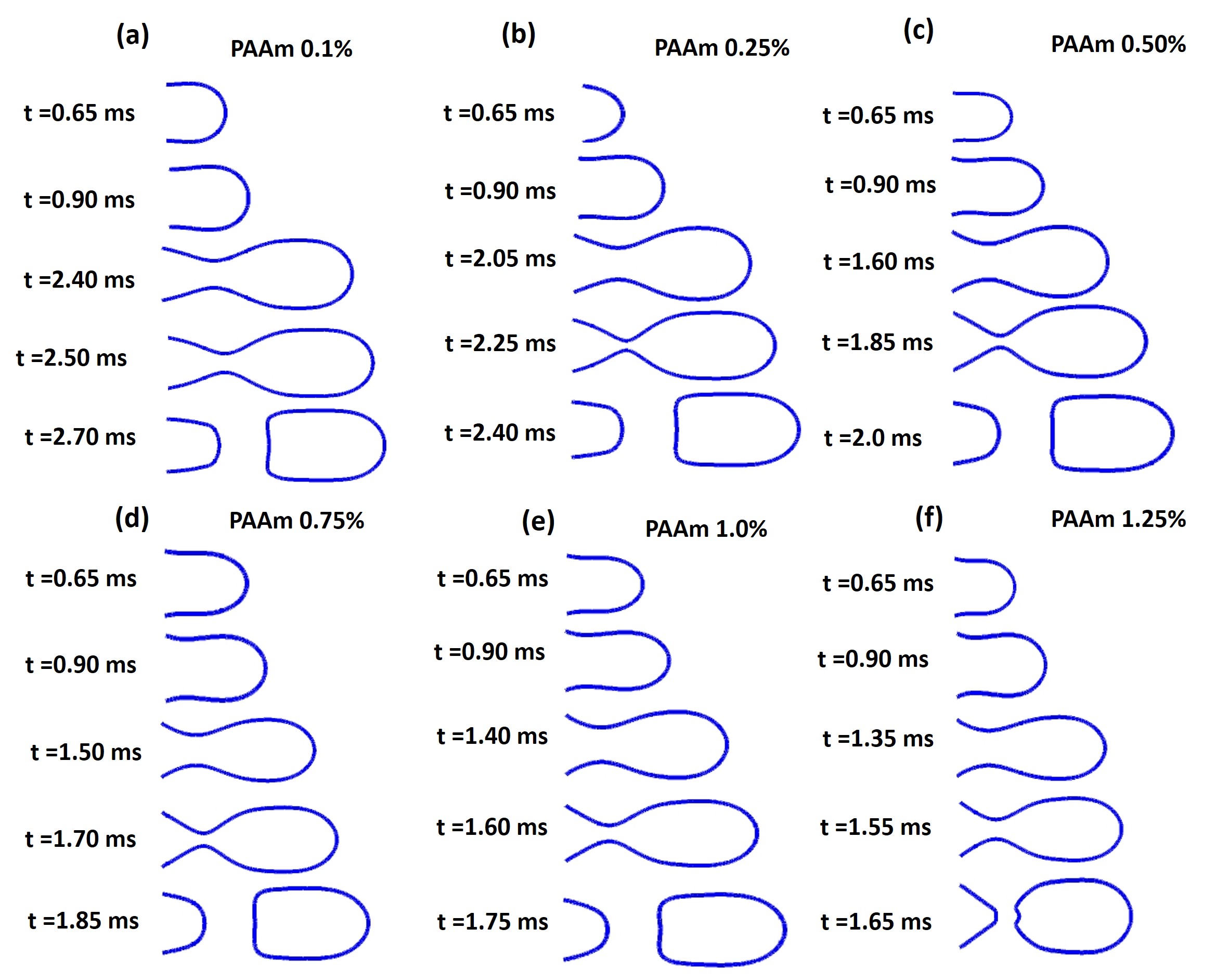}
	\caption{\label{fig:paam_12} Taylor bubble evolution in a co\textendash flow microchannel having different PAAm solution for (a) PAAm 0.1 \%, (b) PAAm 0.25\%, (c) PAAm 0.50\%, (d) PAAm 0.75 \%, (e) PAAm 1.0\%, and (f) PAAm 1.25 \%  at $U_{L}$ = 0.5 m/s and $U_{G}$ = 0.5 m/s.}
\end{figure}

The dynamics of Taylor bubble pinch\textendash off in different PAAm solutions are also investigated by analyzing six sequential images obtained from our simulation, as shown in Fig.~\ref{fig:paam_12}.
With increasing PAAm concentration, the final shape of the bubble changes (Fig.~\ref{fig:paam_12}a-f) and the bubble pinch\textendash off process accelerates with increasing viscous nature of the continuous liquid phase. This is ascribed to increased viscous drag on the gas phase resulting from increasing viscosity and shear stress at higher PAAm concentration, which dominates over interfacial and inertia forces. This also causes relatively shorter thread before pinch\textendash off for higher PAAm concentration solutions. Rheological properties of the continuous phase are found to influence the leading edge curvature of the Taylor bubble that is essentially determined by the balance between interfacial and viscous forces. With increasing PAAm concentration, Taylor bubble nose curvature decreases due to higher viscous stress on the bubble. 

\begin{figure}[!ht]
	\centering
	\includegraphics[width=0.5\linewidth]{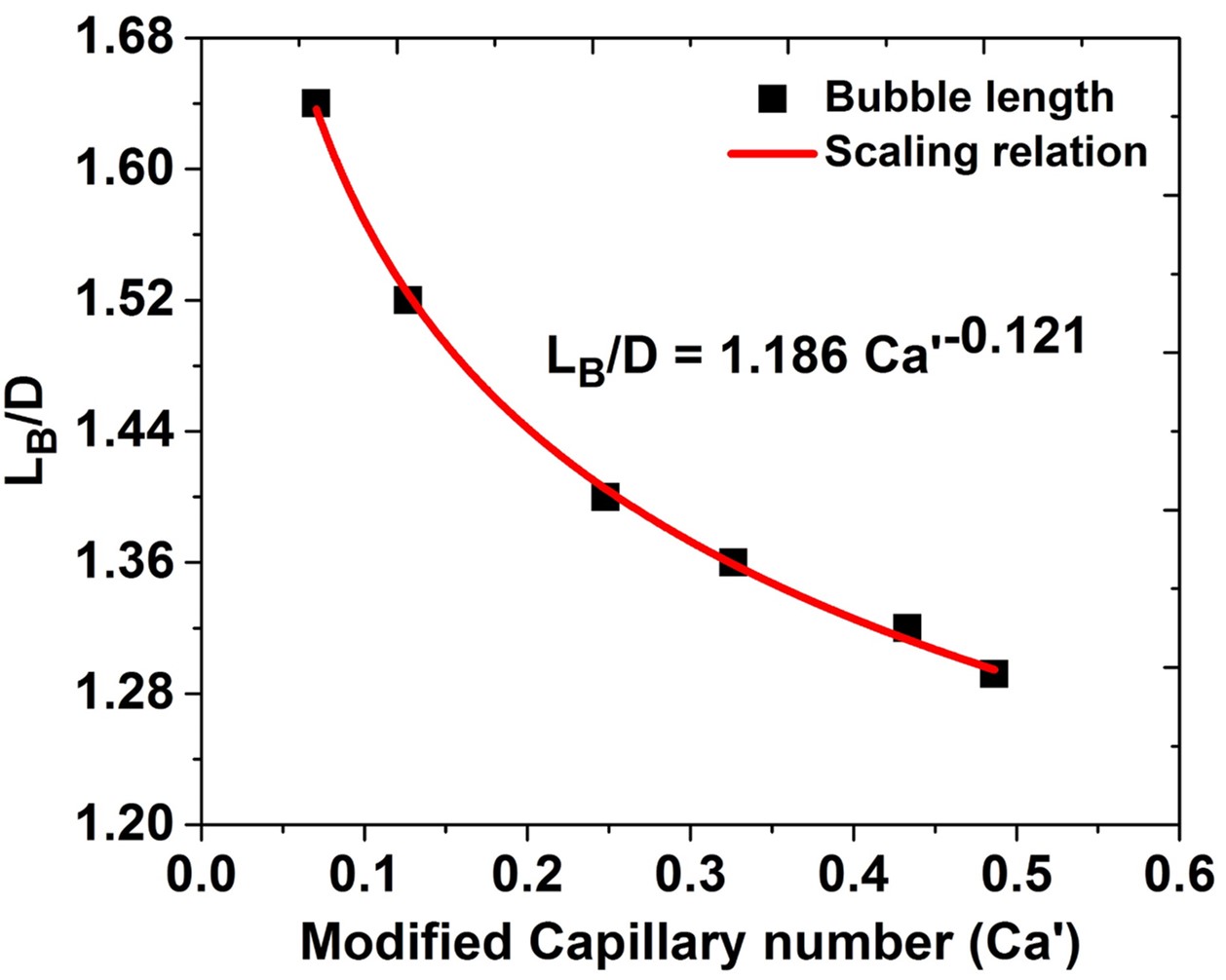}
	\caption{\label{fig:C1} The scaling relation of non\textendash dimensional bubble length with the modified Capillary number ($Ca^{'}$) for different PAAm concentration solutions at $U_{L}$ = 0.5 m/s and $U_{G}$ = 0.5 m/s.}
\end{figure}

Furthermore, dimensionless bubble length ($L_{B}/D$) in all PAAm solutions are plotted as a function of modified Capillary number ($Ca^{'}=KU_B^nD^{(1-n)}/\sigma$) in Fig.~\ref{fig:C1} to determine a scaling law for power-law liquids.
The derived relation for different concentrations of PAAm (0.1\%--1.25\%) shows a maximum deviation of 0.5\%, and is similar in nature to previously reported results \citep{fu-2011,fu-2012b,bai2016dro} for bubble/droplet formation in Newtonian and non\textendash Newtonian liquids in various microchannels, but with different prefactor and exponent. 
\begin{figure}[!ht]
	\centering
	\includegraphics[width=0.8\linewidth]{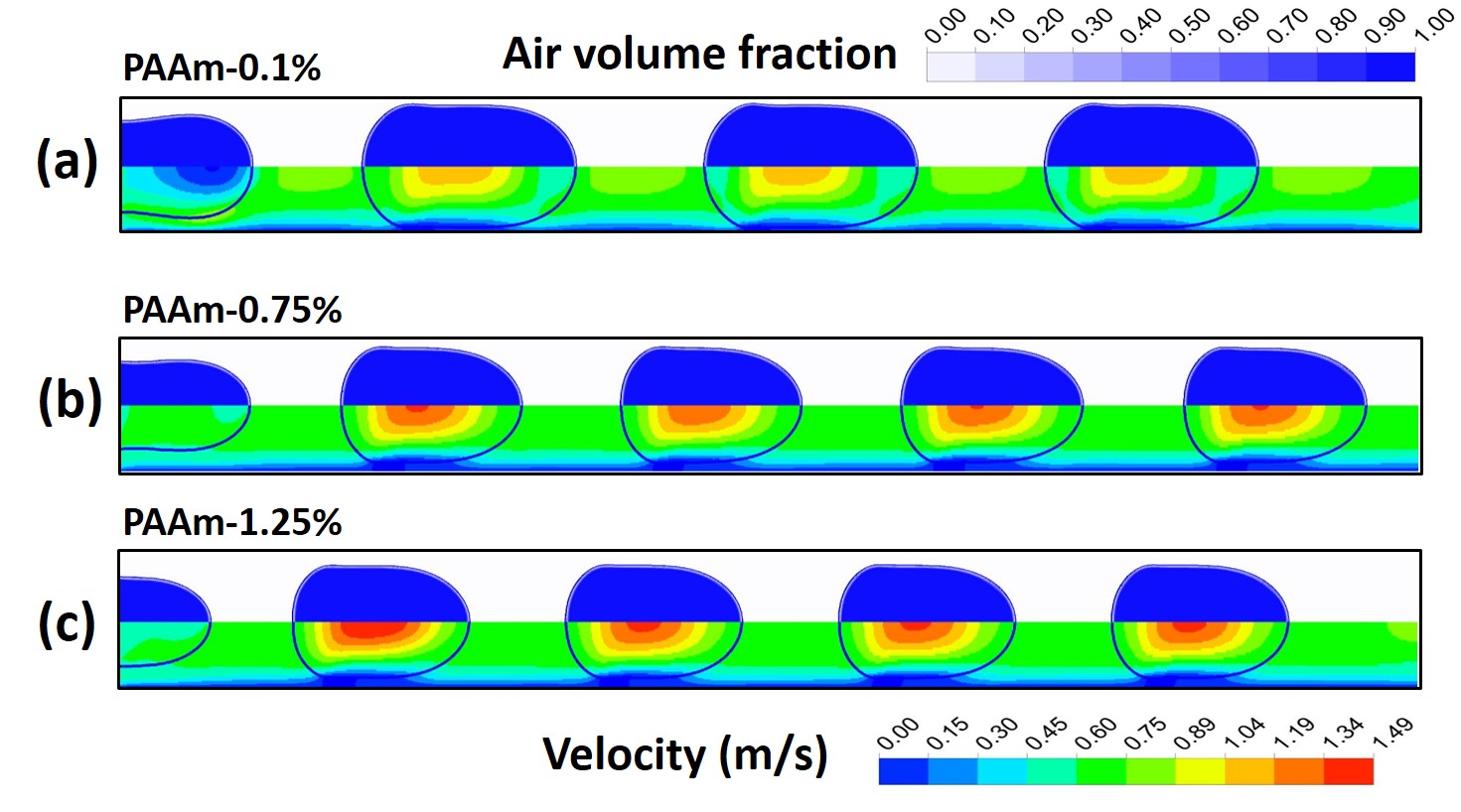}
	\caption{\label{fig:paam_141} Effect of PAAm concentration on velocity distribution (upper halves are volume fraction and lower halves are velocity field) for (a) PAAm\textendash 0.1 wt\%, (b) PAAm\textendash 0.75 wt\%, and (c) PAAm\textendash 1.25 wt\% at $U_{L}$ = 0.5 m/s and $U_{G}$ = 0.5 m/s.}
\end{figure}

\begin{figure}[!ht]
	\centering
	\includegraphics[width=0.8\linewidth]{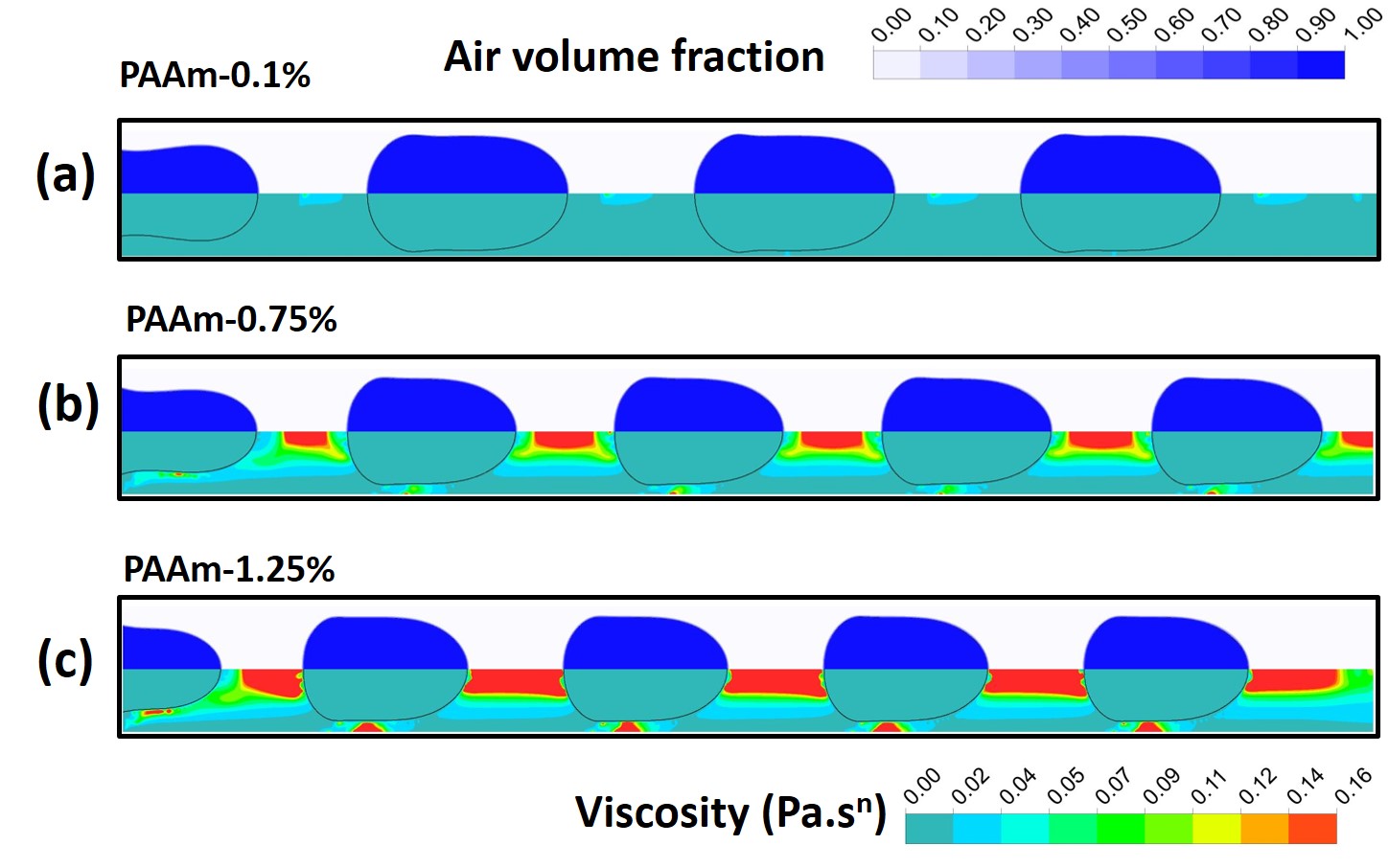}
	\caption{\label{fig:viscosity123} Effect of PAAm concentration on non-homogeneous viscosity distribution (upper halves are volume fraction and lower halves are  on non-homogeneous viscosity) for (a) PAAm\textendash 0.1 wt\%, (b) PAAm\textendash 0.75 wt\%, and (c) PAAm\textendash 1.25 wt\% at $U_{L}$ = 0.5 m/s and $U_{G}$ = 0.5 m/s.}
\end{figure}
To realize the impact of power-law liquid, velocity field inside the bubble is further analyzed for three different PAAm solutions. Fig.~\ref{fig:paam_141} shows the contours of gas phase and velocity fields for various PAAm solutions, which further substantiate the observation of increasing bubble velocity at higher PAAm concentration, as shown in Fig.~\ref{fig:paam_11}c. In all cases, velocity is found to be maximum in the middle of the channel, and its magnitude decreases towards the channel wall, as shown in Fig.~\ref{fig:paam_141}. It can be inferred from the velocity field analysis that with increasing PAAm concentration, the velocity inside bubble increases due to enhanced viscous stress acting on the Taylor bubble resulting from higher effective viscosity and liquid film thickness around the bubble. 
\begin{figure}[!ht]
	\centering
	\includegraphics[width=\linewidth]{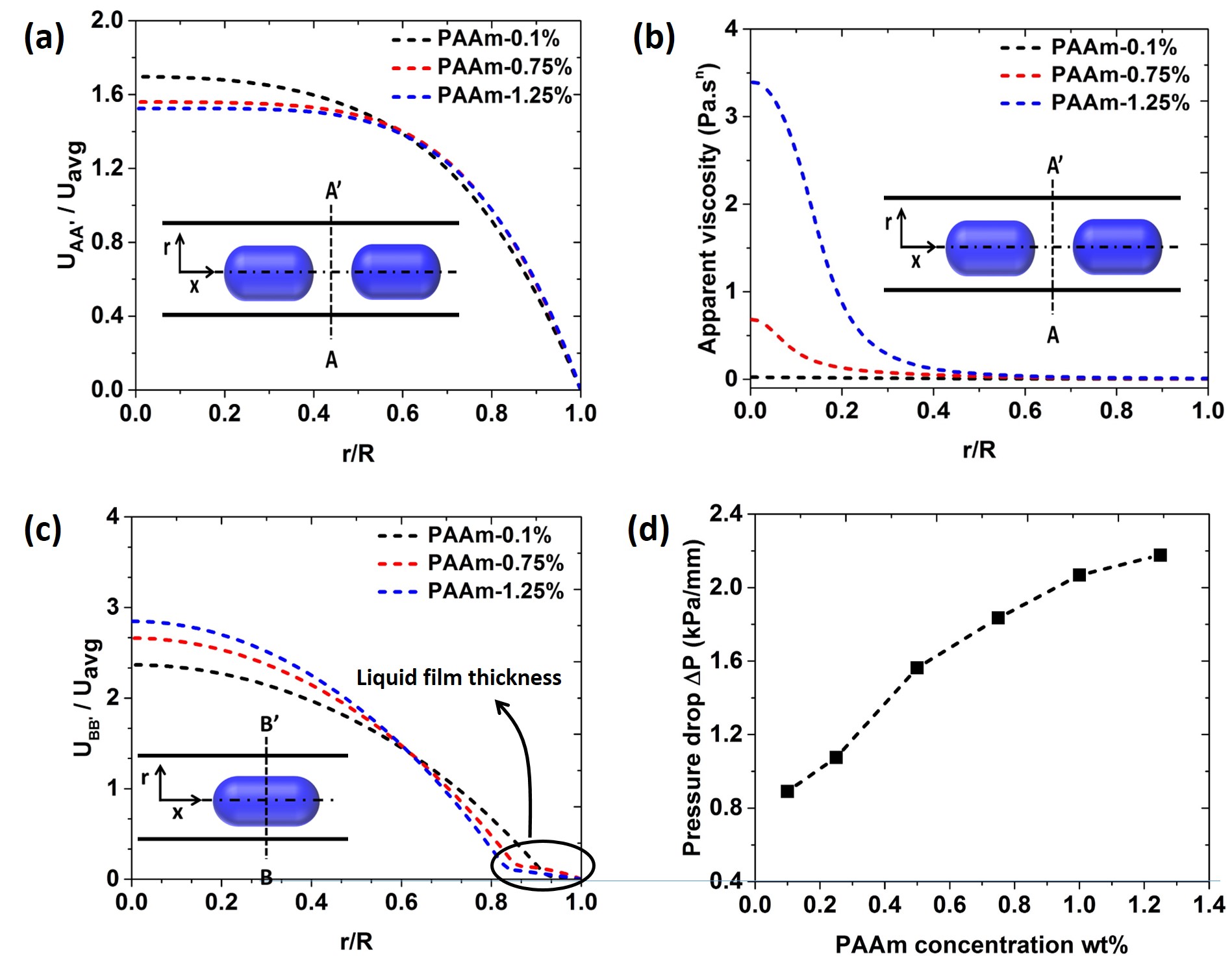}
	\caption{\label{fig:paam_14} Effect of PAAm concentration on (a) velocity profiles in the middle of the liquid slug, (b) effective viscosity along the channel radius in the middle of the liquid slug, (c) velocity profile in the middle of the Taylor bubble, and (d) pressure drop at $U_{L}$ = 0.5 m/s and $U_{G}$ = 0.5 m/s.}
\end{figure}
Fig.~\ref{fig:viscosity123} illustrates non-homogeneous viscosity distribution along with volume fraction for various PAAm solutions in the microchannel. It is noticeable that the heterogeneity in viscosity distribution increases with increasing effective viscosity, which eventually depends on PAAm concentration (Fig.~\ref{fig:paam_11}a). This distribution is also closely related to velocity field around the bubble, as shown in Fig.~\ref{fig:paam_141}. Figs.~\ref{fig:viscosity123}a-c show that with increasing PAAm concentration, maximum magnitude of effective viscosity around the Taylor bubble considerably increases, which is in agreement with the experimental observation of \citet{roumpea2017experimental}, and \citet{fu-2012b} obtained by $\mu$PIV. This effect of fluid rheology becomes discernible when the velocity profiles in three different PAAm solutions are analyzed for both liquid slug and Taylor bubble. From Fig.~\ref{fig:paam_14}a, it is evident that the centerline velocity in the liquid slug significantly changes with an increase in PAAm concentration. In higher PAAm concentrations, flattened profiles are observed. At lower concentration (i.e., PAAm\textendash0.1 wt\%), higher velocity in the center of the microchannel is observed (Fig.~\ref{fig:paam_14}a). The non\textendash homogeneous viscosity distribution in the middle of the liquid slug along the channel radius shows that the viscosity of the PAAm solutions significantly increases with concentration in the center of the microchannel (Fig.~\ref{fig:paam_14}b). Fig.~\ref{fig:paam_14}c quantitatively depicts the effect of PAAm concentration on the velocity inside the Taylor bubble, as discussed earlier. Fig.~\ref{fig:paam_14}d shows the influence of PAAm concentration on the pressure drop of the system, which increases with concentration due to the increase in viscous effects of the liquid phase. 

  
\subsection{Effect of inlet velocity}

\begin{figure}[h]
	\centering
	\includegraphics[width=\linewidth]{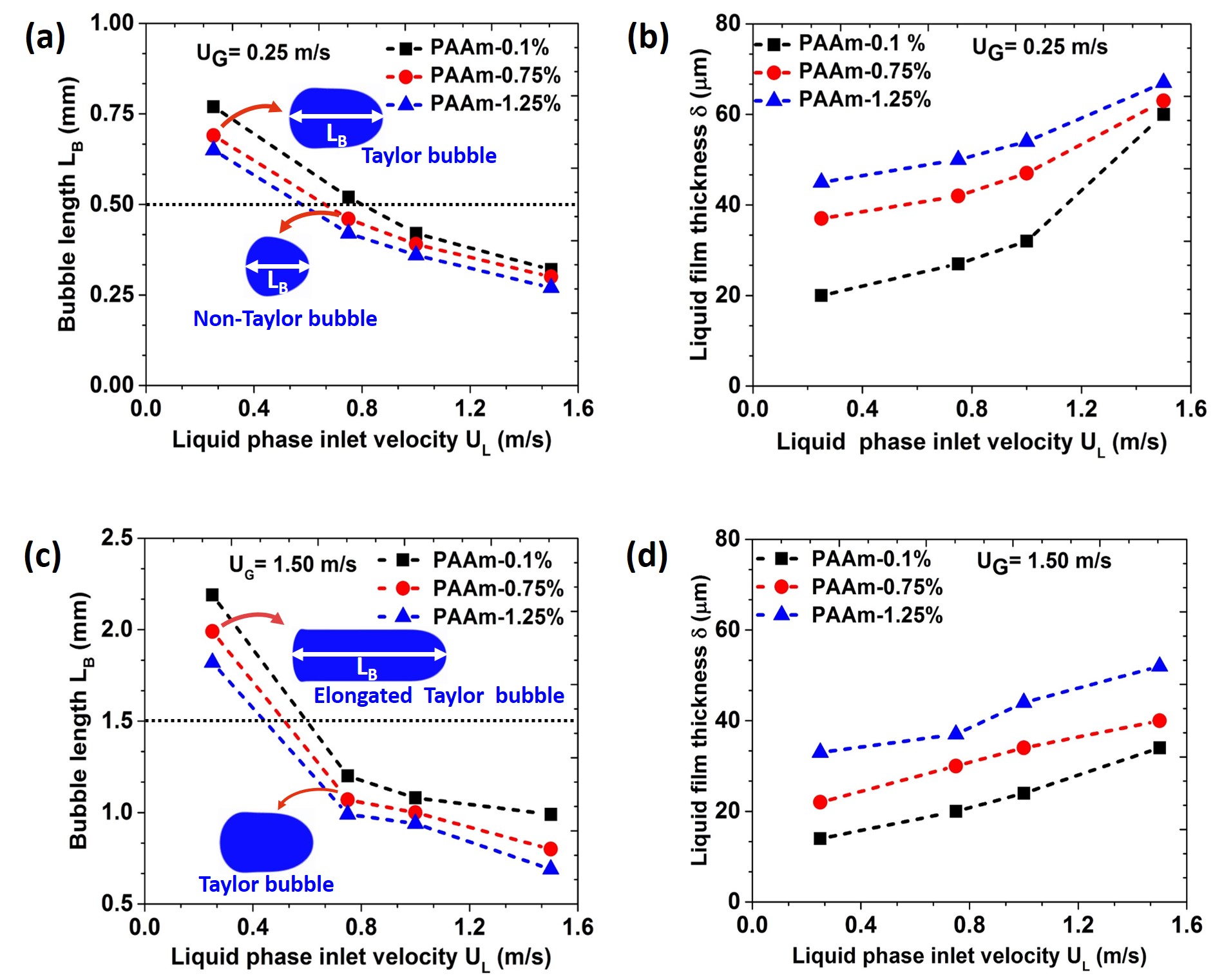}
	\caption{\label{fig:LSP1}Effect of liquid inlet velocity on (a) Taylor bubble length at $U_G$=0.25 m/s, (b) liquid film thickness at $U_G$=0.25 m/s, (c) Taylor bubble length at $U_G$=1.50 m/s, and (d) liquid film thickness at $U_G$=1.50 m/s.}
\end{figure}

\begin{figure}[h]
	\centering
	\includegraphics[width=0.9\textwidth]{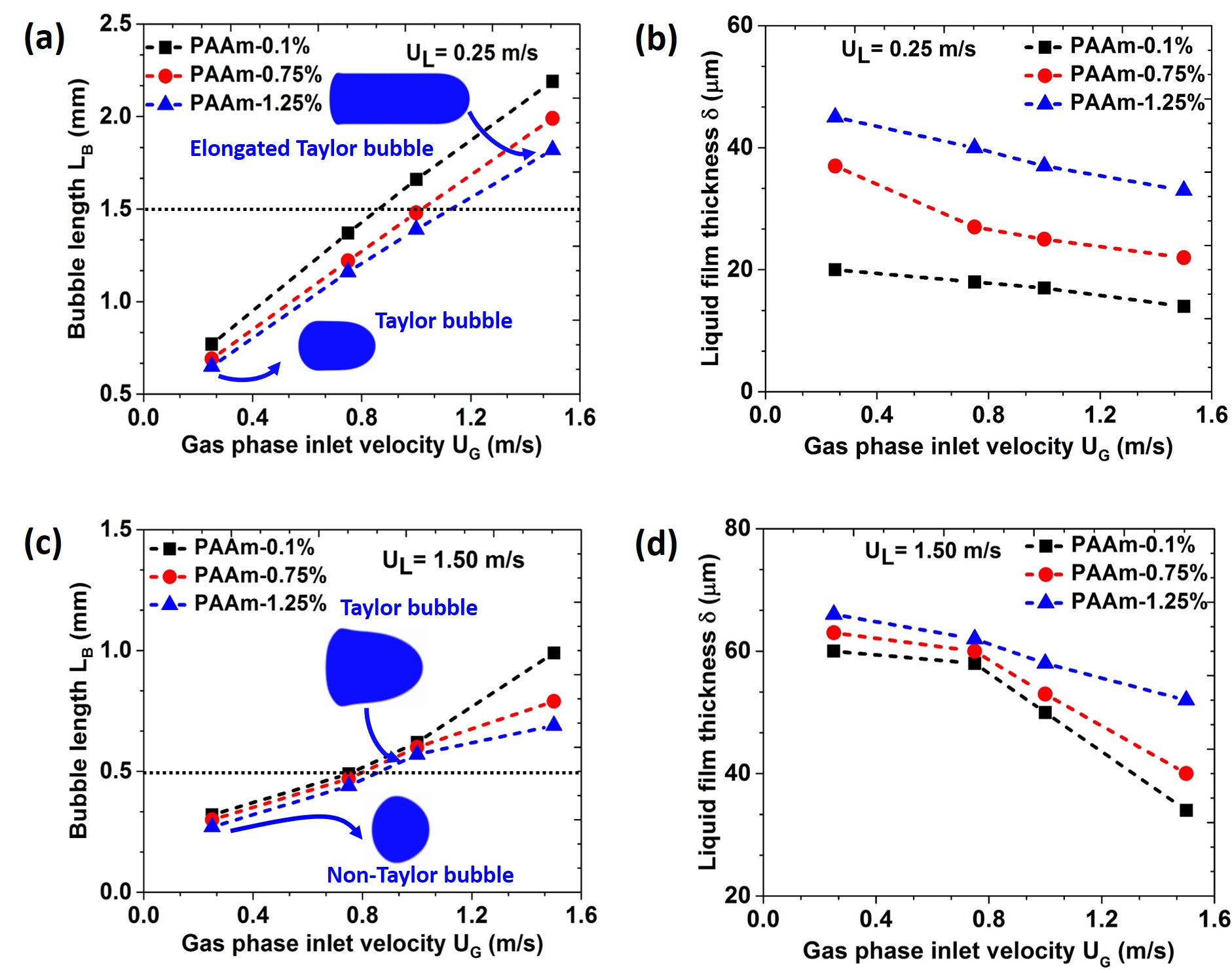}
	\caption{\label{fig:GSP2} Effect of gas inlet velocity on (a) Taylor bubble length at $U_L$=0.25 m/s, (b) liquid film thickness at $U_L$=0.25 m/s, (c) Taylor bubble length at $U_L$=1.50 m/s, and (d) liquid film thickness at $U_L$=1.50 m/s.}
\end{figure}

The effect of continuous phase velocity on bubble length is described in Fig.~\ref{fig:LSP1} for three different PAAm solutions. Figs.~\ref{fig:LSP1}a and \ref{fig:LSP1}c demonstrate that the bubble length decreases with increasing continuous phase velocity for two different gas inlet velocities ($U_G=0.25$ m/s and $U_G=1.50$ m/s, respectively) and fixed liquid properties. 
This is attributed to higher inertia force imparted on the gas phase. However, the liquid film thickness increases with increasing liquid inlet velocity, as shown in Figs.~\ref{fig:LSP1}b and \ref{fig:LSP1}d. In this work, the bubble shape is broadly categorized into three types based on its length ($L_B$) with respect to capillary diameter ($D=0.5~mm$) such as, non\textendash Taylor bubble ($L_B$ \textless D), Taylor bubble ($L_B$ \textgreater D) and elongated Taylor bubble ($L_B>3~D$). In the studied range of $U_L$ (0.25\textendash 1.50 m/s) and PAAm concentrations, Fig.~\ref{fig:LSP1}a also depicts two distinct shapes of the bubbles, which are observed with increasing $U_L$. For $U_G=0.25$ m/s and at lower liquid velocity, Taylor bubbles ($L_B>D$) are formed however, non\textendash Taylor bubbles ($L_B$ \textless D) are detected at higher liquid velocity. Furthermore, at a higher gas inlet velocity ($U_G$=1.50 m/s) but at lower $U_L$, elongated Taylor bubbles ($L_B>3~D$) are observed, as shown in Fig.~\ref{fig:LSP1}c. The effect of gas inlet velocity is also investigated by keeping other process conditions unaltered. Fig.~\ref{fig:GSP2} depicts that on increasing the gas inlet velocity, bubble length increases (Fig.~\ref{fig:GSP2}a and Fig.~\ref{fig:GSP2}c), and the surrounding liquid film thickness decreases (Fig.~\ref{fig:GSP2}b and Fig.~\ref{fig:GSP2}d). 
At relatively lower liquid inlet velocity ($U_L$=0.25 m/s), flow transition from Taylor bubble to elongated Taylor bubble is detected with increasing gas inlet velocity (Fig.~\ref{fig:GSP2}a). However, at higher liquid inlet velocity ($U_L$ =1.50 m/s), the transition occurs from non-Taylor bubble to Taylor bubble with increasing gas velocity (Fig.~\ref{fig:GSP2}c). 

\begin{figure}[h]
	\centering
	\includegraphics[width=0.65\linewidth]{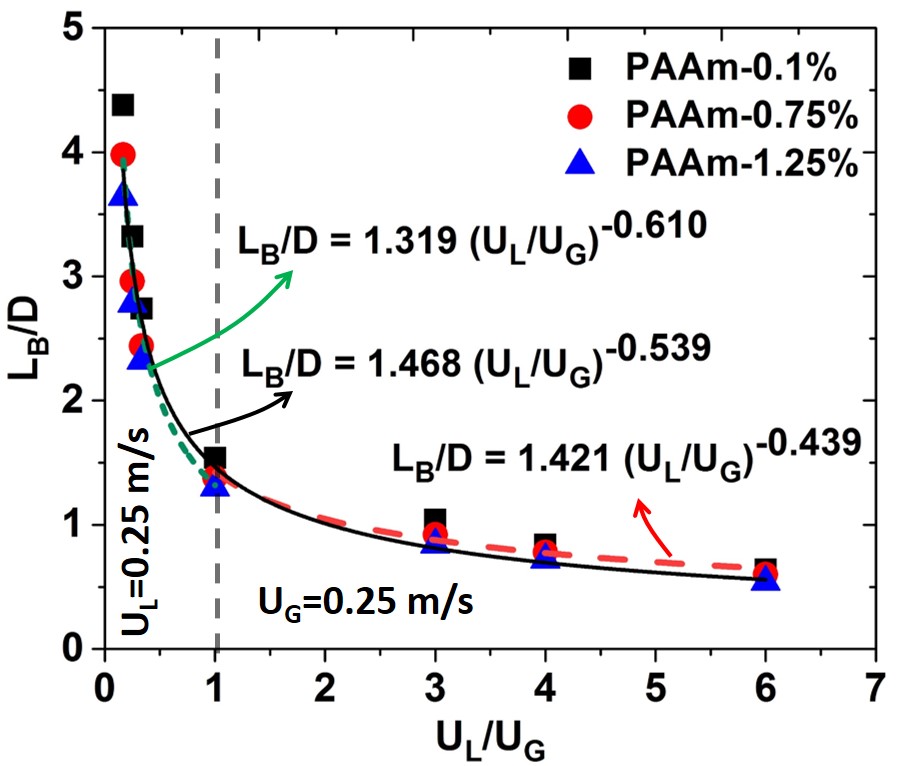}
	\caption{\label{fig:CR1} The scaling of non-dimensional bubble length with fluid inlet velocity ratio ($U_{L}/U_{G}$) for PAAm concentration liquids. Dotted line at $U_{L}/U_{G}=1$ demarcates the region of constant velocity of one fluid phase.}
\end{figure}

Fig.~\ref{fig:CR1} shows the scaling of non-dimensional bubble length with $U_L/U_G$ ratio, where either gas ($U_G$=0.25 m/s) or liquid ($U_L$=0.25 m/s) velocity was kept constant. The scaling relation for the variation of $U_L/U_G$ from 0 to 1 indicates the influence of gas inlet velocity when liquid inlet velocity was fixed at 0.25 m/s. Rest of $U_L/U_G$ values depict influence of liquid inlet velocity, which are already discussed in Figs.~\ref{fig:LSP1}a and \ref{fig:GSP2}a.
For $U_L$ = 0.25 m/s, the proposed relation $L_B/D=1.319(U_L/U_G)^{-0.610}$ provides result with a maximum deviation of 14\% for any PAAm concentration. However, the scaling relation for $U_G$ = 0.25 m/s, $L_B/D=1.421(U_L/U_G)^{-0.439}$ indicates a maximum variation of 19\% in any case within the range of studied conditions. To encompass the complete range of inlet velocities by a single relation, $L_B/D=1.468(U_L/U_G)^{-0.539}$, a maximum deviation of 30\% is observed for the considered range of PAAm concentrations.     
   

\subsection{Flow regimes maps}

Knowledge of flow patterns forming under given inlet and operating conditions is essential for understanding the behavior of gas\textendash liquid microsystems. Nonetheless, it is cumbersome to define a flow pattern map that includes the influence of all plausible parameters affecting the transition. The flow maps reported in the literature are generally proposed for Newtonian fluids. Here, the flow patterns that form under different velocities are reported for two PAAm solutions (0.10 wt\% and 1.25 wt\%). 
\begin{figure}[h]
	\centering
	\includegraphics[width=\linewidth]{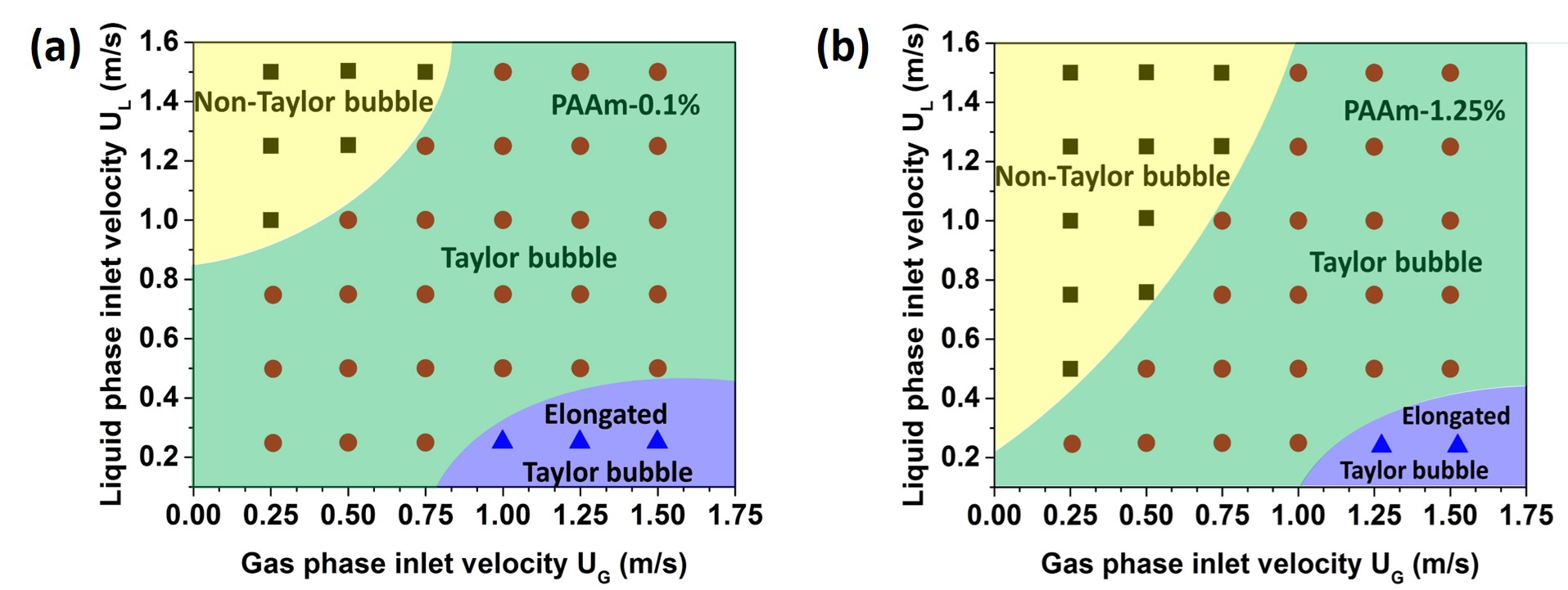}
	\caption{\label{fig:Regime} Flow regime map of Taylor bubble formation as a function of $U_G$ and $U_L$ for air-PAAm system in a co-flow microchannel with (a) PAAm 0.10 wt\%, and (b) PAAm 1.25 wt\%.}
\end{figure}
The flow regime maps of air\textendash PAAm solution in the co-flow microchannel are shown in Fig.~\ref{fig:Regime} with gas ($U_G$) and liquid ($U_L$) inlet velocities as coordinates. Three types of bubble shape classification that were mentioned earlier, non-Taylor bubble, Taylor bubble, and elongated Taylor bubble, are shown in the flow maps (Figs.~\ref{fig:Regime}a and \ref{fig:Regime}b). It can be noted from Figs.~\ref{fig:Regime}a and \ref{fig:Regime}b that increasing liquid flow rate leads to disruption of Taylor bubbles ($L_B > D$) toward bubbly flow ($L_B < D$), and the phenomenon is more pronounced in relatively higher concentration of PAAm solution (1.25 wt\%) as compared to PAAm-0.10 wt\% solution, which is nearly Newtonian in nature. In case of higher PAAm concentration (PAAm-1.25 wt\%), elongated Taylor bubble regime occupies small area compared to lower concentration solutions.\\
\begin{figure}[h]
	\centering
	\includegraphics[width=0.9\textwidth]{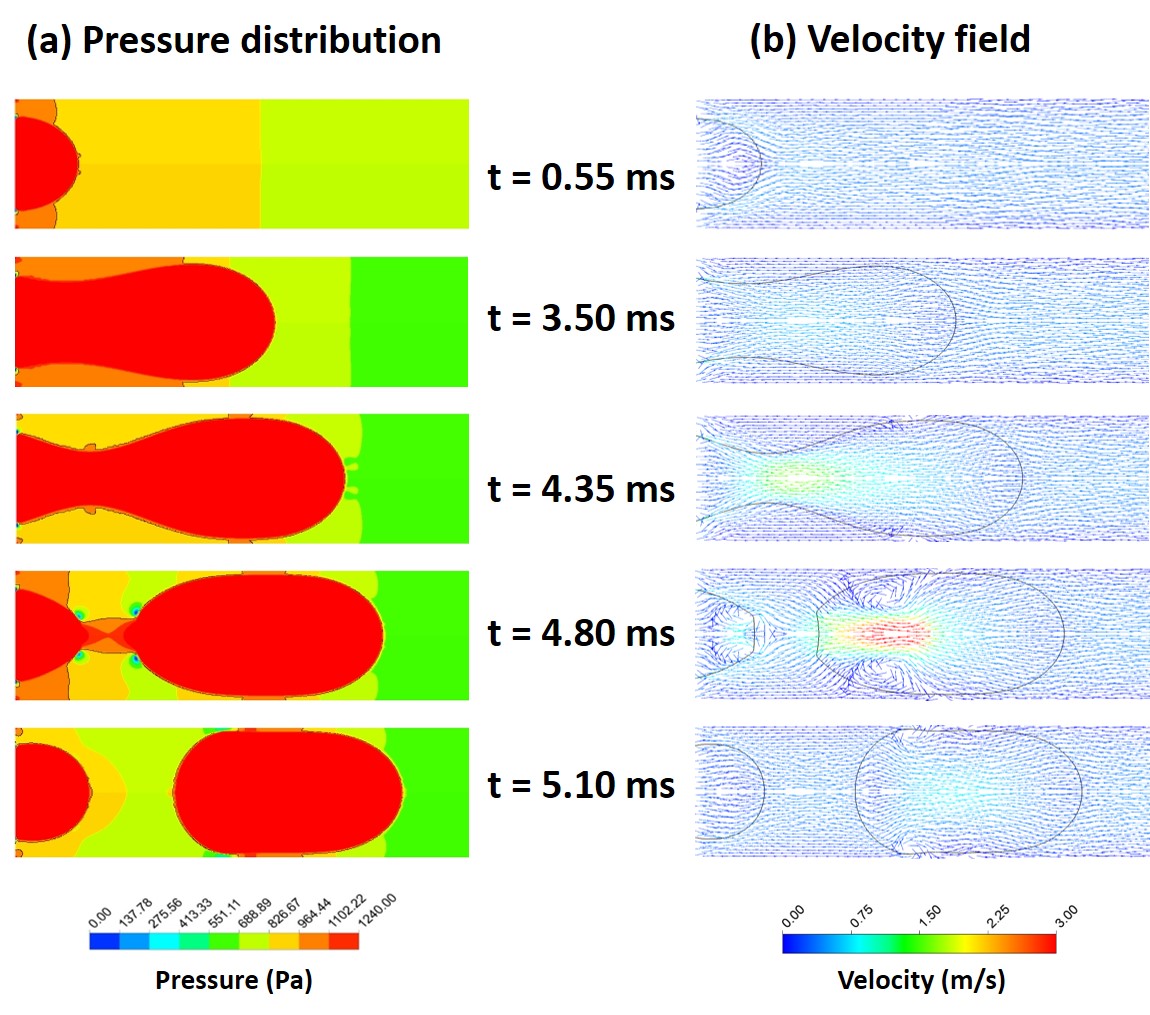}
	\caption{\label{fig:Regime1}(a)Pressure and (b) velocity field distribution in the dripping regime for air-PAAm 0.1 wt\% system at $U_G$=0.25 m/s and $U_L$=0.25 m/s.}
\end{figure}
Fig.~\ref{fig:Regime1} demonstrates pressure and velocity field distributions in the dripping regime ($U_G$=0.25 m/s, and $U_L$=0.25 m/s). Sequential images of pressure (Fig.~\ref{fig:Regime1}a) and velocity field (Fig.~\ref{fig:Regime1}b) evolution mainly consists of two stages; bubble growth ($t=0.55~ms - 4.35~ms$) and pinch-off ($t=4.35~ms - 5.10~ms$). It is apparent from Fig.~\ref{fig:Regime1}a that the pressure distribution inside the Taylor bubble remains constant however, it decreases around the gas phase as the bubble grows in size. A pressure gradient develops from the gas inlet to the bubble neck and it leads to bubble pinch-off, as illustrated in Fig.~\ref{fig:Regime1}a at $t=4.80~ms$. With the onset of pressure gradient, velocity recirculation is observed (Fig.~\ref{fig:Regime1}b at $t= 4.35~ms$), which results in enhanced shear rate until the bubble snaps off.


\begin{figure}[!ht]
	\centering
	\includegraphics[width=0.9\textwidth]{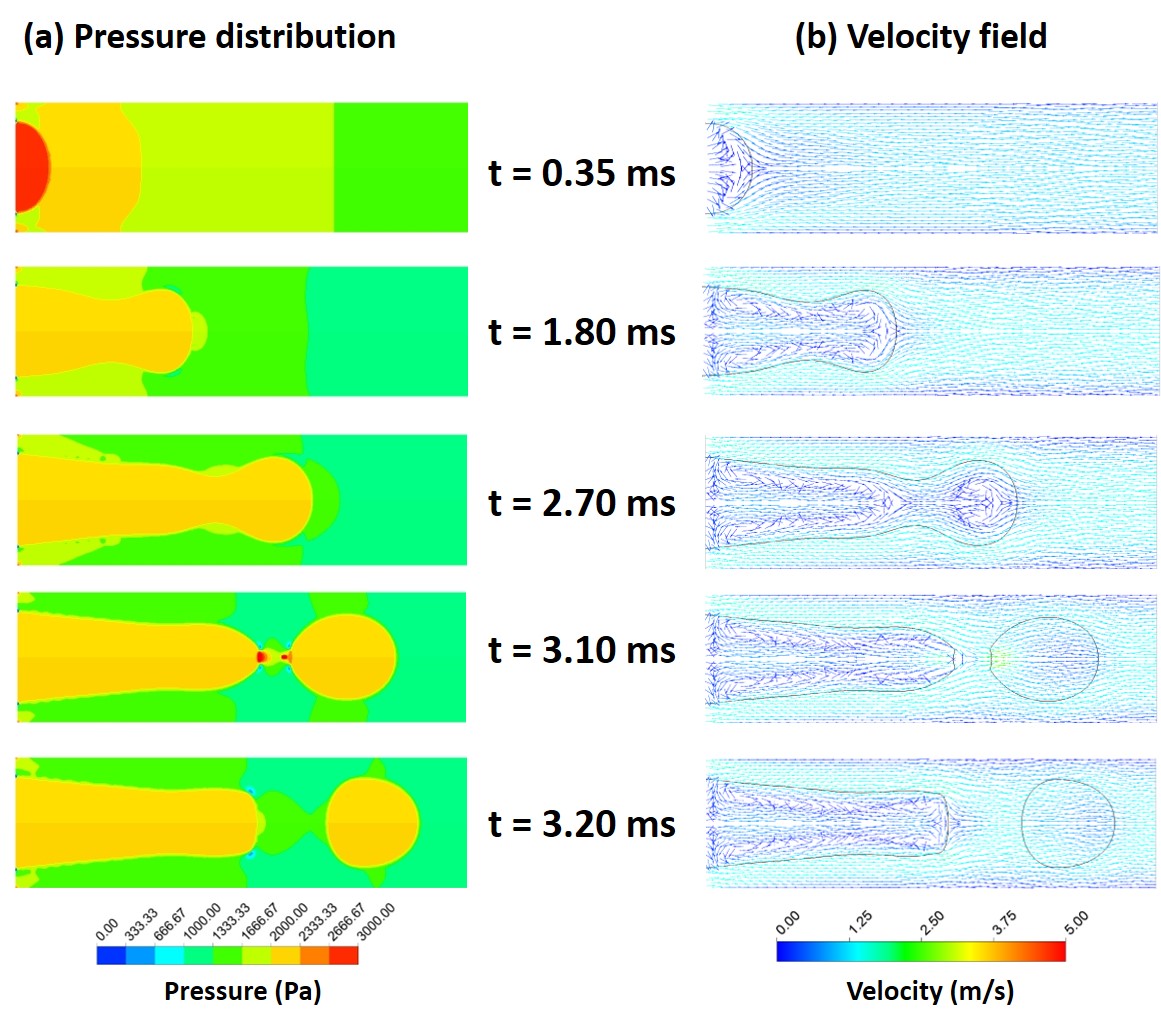}
	\caption{\label{fig:Regime31} (a) Pressure and (b) velocity field distribution in the jetting regime for air-PAAm 0.1 wt\% system at $U_G$=0.25 m/s and  $U_L$=1.5 m/s.}
\end{figure}
With increasing liquid inlet velocity from $U_L$=0.25 m/s to $U_L$=1.5 m/s, flow regime is observed to shift from dripping to jetting. Fig.~\ref{fig:Regime31} portrays pressure and velocity field distributions in the jetting regime, where the thread length (from inlet to the position, where bubbles snap-off) increases due to growing inertial force along the axial direction up to a critical limit. This phenomenon can be observed from Fig.~\ref{fig:Regime31}a at $t=3.10~ms$, and corroborates with the experimental realization of \citet{deng2017} and \citet{lan2015numerical}. Alike dripping regime, pressure distribution around the gaseous thread changes with the bubble formation however, the magnitude is significantly higher in jetting regime. A high pressure point is observed at the nose of gas phase thread after the detachment of bubble, as shown in Fig.~\ref{fig:Regime31}a at $t=3.10~ms$. Fig.~\ref{fig:Regime31}b indicates considerable velocity recirculation inside the gaseous thread as compared to the dripping regime.

\subsection{Effect of surface tension}
Taylor bubble flow in microchannels has strong dependence on the surface tension and viscous forces. Control of surface tension in aqueous solution is an active function of surfactant concentration. Typically, the surface tension of solution decreases with increasing the surfactant concentration. Several researchers have reported the influence of surface tension for air\textendash water system using different concentrations of sodium dodecyl sulfate (SDS) \citep{hernainz2002,xu2006,dietri-2008}. In this work, the surface tension of three different aqueous solutions of PAAm is varied from their respective reference values (see Table~\ref{tab:power_law_data_picchi}) to a notional value of 40 mN/m, which is close to the experimental data for air\textendash Water+SDS 0.15 wt\% system \citep{dietri-2008}.
\begin{figure}[h]
	\centering
	\includegraphics[width=\linewidth]{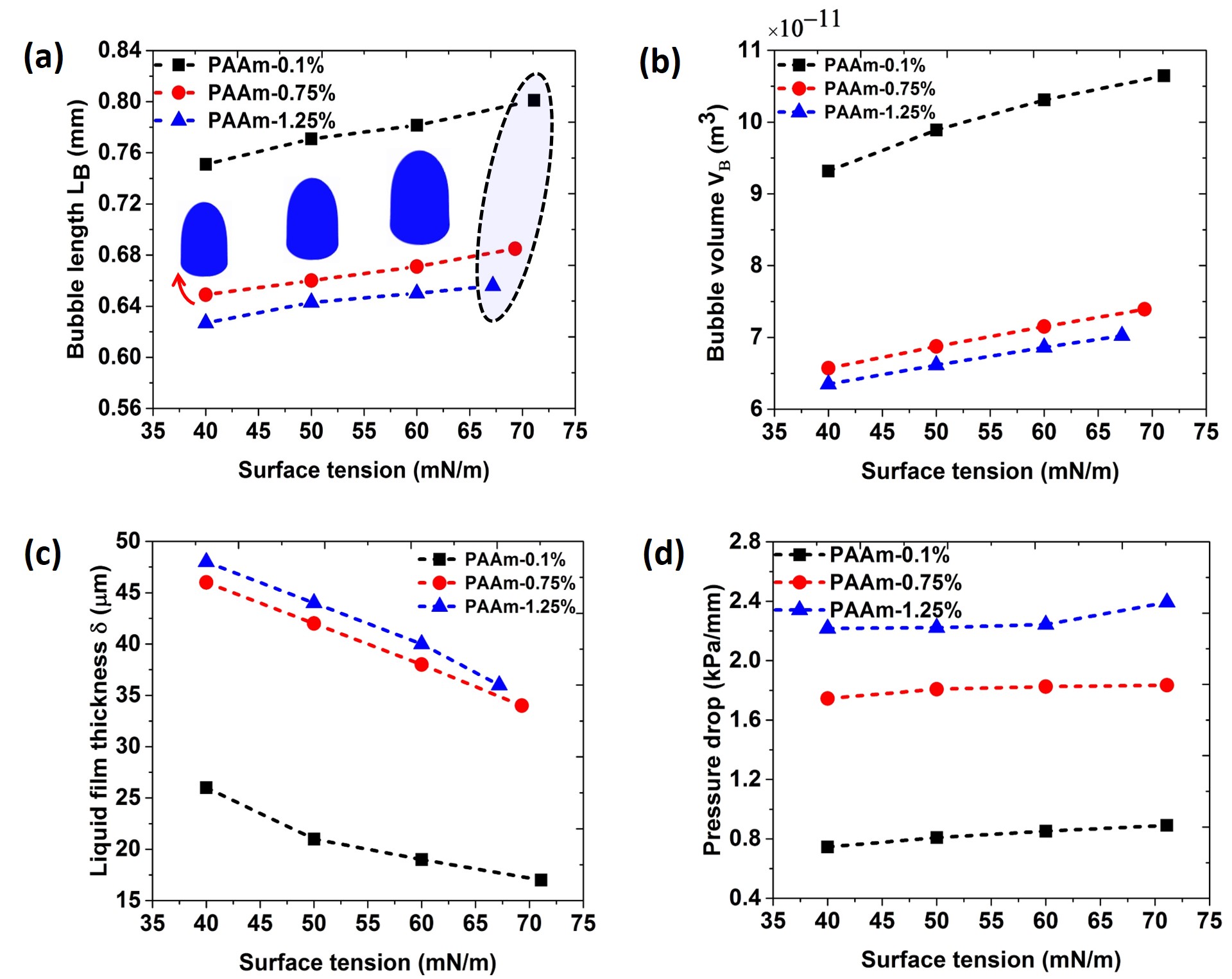}
	\caption{\label{fig:Surfacetension_1} Effect of surface tension on (a) Taylor bubble length (reference values of surface tension are encircled), (b) bubble volume, (c) liquid film thickness, and (d) overall pressure drop for different PAAm solutions.}
\end{figure}
Fig.~\ref{fig:Surfacetension_1}a shows that with increasing surface tension, the bubble length increases and the shape changes due to higher surface tension force in all PAAm solutions. The bubble detachment process accelerates at lower surface tension resulting smaller bubble volume, as shown in Fig.~\ref{fig:Surfacetension_1}b. In line with discussion in the previous section, surrounding liquid film thickness decreases as the bubble volume increases with increasing surface tension (Fig.~\ref{fig:Surfacetension_1}c). Additionally, the overall pressure drop in the microchannel is also analyzed and is found to increase with surface tension for all PAAm solutions, as depicted in Fig.~\ref{fig:Surfacetension_1}d.      

\begin{figure}[h]
	\centering
	\includegraphics[width=0.6\linewidth]{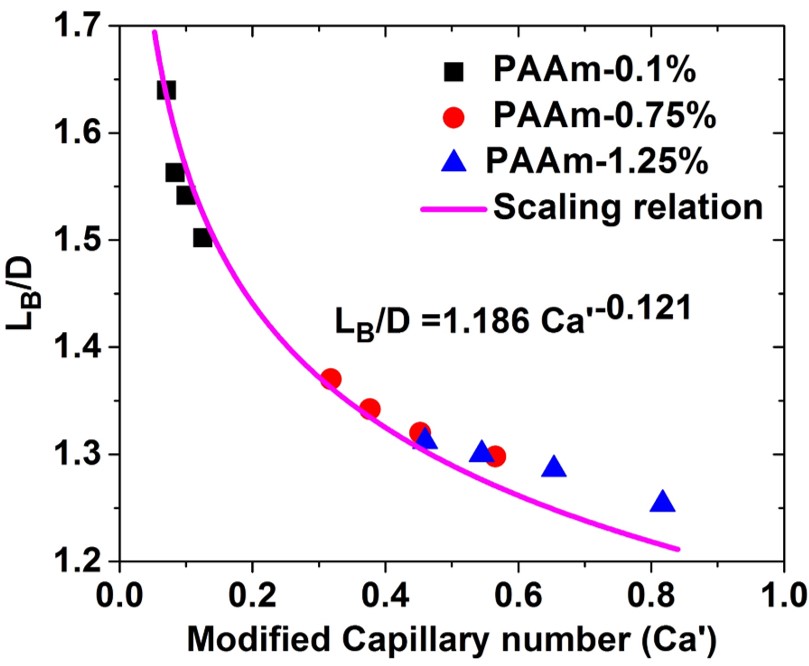}
	\caption{\label{fig:Surfacetension_12} The scaling of non-dimensional bubble length with $Ca^{'}$ at $U_{L}$ = 0.5 m/s and $U_{G}$ = 0.5 m/s for different PAAm solutions.}
\end{figure}

Fig.~\ref{fig:Surfacetension_12} shows that previously derived scaling law (in Fig.~\ref{fig:C1}) fits well (maximum deviation of 3\%) in proposing the dimensionless bubble length as a function of $Ca^{'}$ for different PAAm solutions, even when only surface tension is varied ($0.04\leq \sigma \leq 0.0712$). 


\section{Conclusions}
A systematic computational study on the Taylor bubble formation in aqueous solutions of PAAm with air, as the gaseous phase, is carried out in a circular co-flow microchannel using CLSVOF method. The influences of PAAm concentrations, gas/liquid inlet velocities, and surface tension are methodically explored through the detailed analysis of Taylor bubble length, shape, surrounding liquid film thickness, and pressure drop of the system. Liquid film thickness between the bubble and channel wall is precisely captured to understand its effect on the bubble characteristics. Bubble length is found to decrease with increasing concentration of PAAm and liquid inlet velocity. However, it increases with increasing surface tension and gas phase velocity. Three different types of bubbles are identified, and the flow regime maps for power-law liquids are developed based on gas-liquid inlet velocities. Scaling laws are proposed to determine the bubble length based on gas-liquid velocity ratio, and the modified Capillary number that takes into consideration the continuous phase rheological properties and surface tension of the system. These findings provide better understanding of the Newtonian bubble formation in a non\textendash Newtonian flow system, and can also aid in formulating new guidelines to produce the desired Taylor bubbles.




\section*{Nomenclature}

\noindent $D$ = diameter (m) \\
$U$  =  velocity (m/s) \\
$\hat{N}$	= unit normal vector\\
$p$ = pressure (Pa)\\
$n$ = power-law index\\
$K$ = consistency index ($Pa.s^{n}$)\\
$t$ = flow time (s)\\
$\bigtriangleup$P =  pressure drop (Pa)\\
$\textit{H}(\varphi)$ = Heaviside function \\
$\vec{\chi}$ = position vector \\
$\textit{a}$ =  interface thickness (m)\\
$\textit{d}$ =  shortest distance \\
$\delta ( \varphi )$ =  Direct delta function \\ 
\textit{Greek symbols}\\	
$\alpha$  = volume fraction\\
$\dot{\gamma } $ = shear rate (1/s)\\
$\delta$ = liquid film thickness (m) \\ 
$\theta$ = contact angle (\textdegree)\\
$\kappa_{n}$ = radius of curvature \\
$\eta$ = dynamic viscosity (kg/m.s)\\
$\rho$  = density ($kg/m^{3}$)\\
$\sigma$  = surface tension (N/m)\\ 
$\overline{\overline\tau}$ = shear stress (Pa)\\
$\varphi $  = level set function \\
\textit{Subscripts}\\
$B$ = bubble \\
\textit{eff} = effective \\
$G$ = gas\\
$L$ = liquid\\

\section*{References}
\bibliography{sekhar_III}

\begin{thebibliography}{80}
\expandafter\ifx\csname natexlab\endcsname\relax\def\natexlab#1{#1}\fi
\providecommand{\url}[1]{\texttt{#1}}
\providecommand{\href}[2]{#2}
\providecommand{\path}[1]{#1}
\providecommand{\DOIprefix}{doi:}
\providecommand{\ArXivprefix}{arXiv:}
\providecommand{\URLprefix}{URL: }
\providecommand{\Pubmedprefix}{pmid:}
\providecommand{\doi}[1]{\href{http://dx.doi.org/#1}{\path{#1}}}
\providecommand{\Pubmed}[1]{\href{pmid:#1}{\path{#1}}}
\providecommand{\bibinfo}[2]{#2}
\ifx\xfnm\relax \def\xfnm[#1]{\unskip,\space#1}\fi
\bibitem[{Taylor(1961)}]{taylor1961}
\bibinfo{author}{G.~Taylor}, \bibinfo{journal}{J. Fluid Mech.} \bibinfo{volume}{10} (\bibinfo{year}{1961}) \bibinfo{pages}{161--165}.
\bibitem[{G{\"u}nther et~al.(2005)G{\"u}nther, Jhunjhunwala, Thalmann, Schmidt, and Jensen}]{gunther-2005}
\bibinfo{author}{A.~G{\"u}nther}, \bibinfo{author}{M.~Jhunjhunwala}, \bibinfo{author}{M.~Thalmann}, \bibinfo{author}{M.~A. Schmidt}, \bibinfo{author}{K.~F. Jensen}, \bibinfo{journal}{Langmuir} \bibinfo{volume}{21} (\bibinfo{year}{2005}) \bibinfo{pages}{1547--1555}.
\bibitem[{Abiev(2013)}]{abiev2013bubbles}
\bibinfo{author}{R.~S. Abiev}, \bibinfo{journal}{Chem. Eng. J.} \bibinfo{volume}{227} (\bibinfo{year}{2013}) \bibinfo{pages}{66--79}.
\bibitem[{Dang et~al.(2013)Dang, Cheney, Qian, Joo, and Min}]{dang2013reactivity}
\bibinfo{author}{T.-D. Dang}, \bibinfo{author}{M.~A. Cheney}, \bibinfo{author}{S.~Qian}, \bibinfo{author}{S.~W. Joo}, \bibinfo{author}{B.-K. Min}, \bibinfo{journal}{J. Ind. Eng. Chem.} \bibinfo{volume}{19} (\bibinfo{year}{2013}) \bibinfo{pages}{1770--1773}.
\bibitem[{Liedtke et~al.(2016)Liedtke, Bornette, Philippe, and de~Bellefon}]{liedtke2016external}
\bibinfo{author}{A.-K. Liedtke}, \bibinfo{author}{F.~Bornette}, \bibinfo{author}{R.~Philippe}, \bibinfo{author}{C.~de~Bellefon}, \bibinfo{journal}{Chem. Eng. J.} \bibinfo{volume}{287} (\bibinfo{year}{2016}) \bibinfo{pages}{92--102}.
\bibitem[{Zeng et~al.(2012)Zeng, Wang, Wang, Zhang, and Zhang}]{zeng2012novel}
\bibinfo{author}{C.~Zeng}, \bibinfo{author}{C.~Wang}, \bibinfo{author}{F.~Wang}, \bibinfo{author}{Y.~Zhang}, \bibinfo{author}{L.~Zhang}, \bibinfo{journal}{Chem. Eng. J.} \bibinfo{volume}{204} (\bibinfo{year}{2012}) \bibinfo{pages}{48--53}.
\bibitem[{Xiao and Kim(2016)}]{xiao2016numerical}
\bibinfo{author}{X.~Xiao}, \bibinfo{author}{C.~N. Kim}, \bibinfo{journal}{J. Ind. Eng. Chem.} \bibinfo{volume}{38} (\bibinfo{year}{2016}) \bibinfo{pages}{23--36}.
\bibitem[{Triplett et~al.(1999)Triplett, Ghiaasiaan, Abdel-Khalik, and Sadowski}]{tripl-1999}
\bibinfo{author}{K.~Triplett}, \bibinfo{author}{S.~Ghiaasiaan}, \bibinfo{author}{S.~Abdel-Khalik}, \bibinfo{author}{D.~Sadowski}, \bibinfo{journal}{Int. J. Multiphase Flow} \bibinfo{volume}{25} (\bibinfo{year}{1999}) \bibinfo{pages}{377--394}.
\bibitem[{Pohorecki et~al.(2008)Pohorecki, Sobieszuk, Kula, Moniuk, Zieli{\'n}ski, Cyga{\'n}ski, and Gawi{\'n}ski}]{pohore-2008}
\bibinfo{author}{R.~Pohorecki}, \bibinfo{author}{P.~Sobieszuk}, \bibinfo{author}{K.~Kula}, \bibinfo{author}{W.~Moniuk}, \bibinfo{author}{M.~Zieli{\'n}ski}, \bibinfo{author}{P.~Cyga{\'n}ski}, \bibinfo{author}{P.~Gawi{\'n}ski}, \bibinfo{journal}{Chem. Eng. J.} \bibinfo{volume}{135} (\bibinfo{year}{2008}) \bibinfo{pages}{S185--S190}.
\bibitem[{Zaloha et~al.(2012)Zaloha, Kristal, Jiricny, V{\"o}lkel, Xuereb, and Aubin}]{zaloha-2012}
\bibinfo{author}{P.~Zaloha}, \bibinfo{author}{J.~Kristal}, \bibinfo{author}{V.~Jiricny}, \bibinfo{author}{N.~V{\"o}lkel}, \bibinfo{author}{C.~Xuereb}, \bibinfo{author}{J.~Aubin}, \bibinfo{journal}{Chem. Eng. Sci.} \bibinfo{volume}{68} (\bibinfo{year}{2012}) \bibinfo{pages}{640--649}.
\bibitem[{Park et~al.(2009)Park, Shin, Oh, Chung, Huh, and Haam}]{park2009development}
\bibinfo{author}{J.-J. Park}, \bibinfo{author}{Y.~Shin}, \bibinfo{author}{J.-H. Oh}, \bibinfo{author}{C.-H. Chung}, \bibinfo{author}{Y.-J. Huh}, \bibinfo{author}{S.~Haam}, \bibinfo{journal}{J. Ind. Eng. Chem.} \bibinfo{volume}{15} (\bibinfo{year}{2009}) \bibinfo{pages}{618--623}.
\bibitem[{Lu et~al.(2015)Lu, Chen, and Wang}]{lu-2015}
\bibinfo{author}{Q.~Lu}, \bibinfo{author}{D.~Chen}, \bibinfo{author}{Q.~Wang}, \bibinfo{journal}{Chem. Eng. Sci.} \bibinfo{volume}{134} (\bibinfo{year}{2015}) \bibinfo{pages}{96--107}.
\bibitem[{Wang(2015)}]{wang2015speed}
\bibinfo{author}{Z.~Wang}, \bibinfo{journal}{Chem. Eng. J.} \bibinfo{volume}{263} (\bibinfo{year}{2015}) \bibinfo{pages}{346--355}.
\bibitem[{Leclerc et~al.(2010)Leclerc, Philippe, Houzelot, Schweich, and De~Bellefon}]{leclerc2010gas}
\bibinfo{author}{A.~Leclerc}, \bibinfo{author}{R.~Philippe}, \bibinfo{author}{V.~Houzelot}, \bibinfo{author}{D.~Schweich}, \bibinfo{author}{C.~De~Bellefon}, \bibinfo{journal}{Chem. Eng. J.} \bibinfo{volume}{165} (\bibinfo{year}{2010}) \bibinfo{pages}{290--300}.
\bibitem[{Peng et~al.(2015)Peng, Xin, Zhang, Yu, and Zhang}]{peng2015}
\bibinfo{author}{D.~Peng}, \bibinfo{author}{F.~Xin}, \bibinfo{author}{L.~Zhang}, \bibinfo{author}{H.~Yu}, \bibinfo{author}{W.~Zhang}, \bibinfo{journal}{Chem. Eng. Sci.} \bibinfo{volume}{128} (\bibinfo{year}{2015}) \bibinfo{pages}{11--20}.
\bibitem[{Ma et~al.(2017)Ma, Fu, Zhang, Zhu, Ma, and Li}]{ma2017breakup}
\bibinfo{author}{R.~Ma}, \bibinfo{author}{T.~Fu}, \bibinfo{author}{Q.~Zhang}, \bibinfo{author}{C.~Zhu}, \bibinfo{author}{Y.~Ma}, \bibinfo{author}{H.~Z. Li}, \bibinfo{journal}{J. Ind. Eng. Chem.} \bibinfo{volume}{54} (\bibinfo{year}{2017}) \bibinfo{pages}{408--420}.
\bibitem[{Lu et~al.(2016)Lu, Fu, Zhu, Ma, and Li}]{lua-2016}
\bibinfo{author}{Y.~Lu}, \bibinfo{author}{T.~Fu}, \bibinfo{author}{C.~Zhu}, \bibinfo{author}{Y.~Ma}, \bibinfo{author}{H.~Z. Li}, \bibinfo{journal}{Chem. Eng. Sci.} \bibinfo{volume}{152} (\bibinfo{year}{2016}) \bibinfo{pages}{516--527}.
\bibitem[{Dang et~al.(2012)Dang, Kim, Kim, and Kim}]{dang2012preparation}
\bibinfo{author}{T.-D. Dang}, \bibinfo{author}{Y.~H. Kim}, \bibinfo{author}{H.~G. Kim}, \bibinfo{author}{G.~M. Kim}, \bibinfo{journal}{J. Ind. Eng. Chem.} \bibinfo{volume}{18} (\bibinfo{year}{2012}) \bibinfo{pages}{1308--1313}.
\bibitem[{Goel and Buwa(2008)}]{goel-2008}
\bibinfo{author}{D.~Goel}, \bibinfo{author}{V.~V. Buwa}, \bibinfo{journal}{Ind. Eng. Chem. Res.} \bibinfo{volume}{48} (\bibinfo{year}{2008}) \bibinfo{pages}{8109--8120}.
\bibitem[{Shao et~al.(2008)Shao, Salman, Gavriilidis, and Angeli}]{shaods-2008}
\bibinfo{author}{N.~Shao}, \bibinfo{author}{W.~Salman}, \bibinfo{author}{A.~Gavriilidis}, \bibinfo{author}{P.~Angeli}, \bibinfo{journal}{Int. J. Heat Fluid Flow} \bibinfo{volume}{29} (\bibinfo{year}{2008}) \bibinfo{pages}{1603--1611}.
\bibitem[{Chen et~al.(2009)Chen, Kulenovic, and Mertz}]{chen-2009}
\bibinfo{author}{Y.~Chen}, \bibinfo{author}{R.~Kulenovic}, \bibinfo{author}{R.~Mertz}, \bibinfo{journal}{Int. J. Therm. Sci.} \bibinfo{volume}{48} (\bibinfo{year}{2009}) \bibinfo{pages}{234 -- 242}.
\bibitem[{Fu and Ma(2015)}]{fu-2015}
\bibinfo{author}{T.~Fu}, \bibinfo{author}{Y.~Ma}, \bibinfo{journal}{Chem. Eng. Sci.} \bibinfo{volume}{135} (\bibinfo{year}{2015}) \bibinfo{pages}{343--372}.
\bibitem[{Fletcher and Haynes(2017)}]{Fletcher2016}
\bibinfo{author}{D.~F. Fletcher}, \bibinfo{author}{B.~S. Haynes}, \bibinfo{journal}{Chem. Eng. Sci.} \bibinfo{volume}{167} (\bibinfo{year}{2017}) \bibinfo{pages}{334--335}.
\bibitem[{Gupta et~al.(2009)Gupta, Fletcher, and Haynes}]{gupta-2009}
\bibinfo{author}{R.~Gupta}, \bibinfo{author}{D.~F. Fletcher}, \bibinfo{author}{B.~S. Haynes}, \bibinfo{journal}{Chem. Eng. Sci.} \bibinfo{volume}{64} (\bibinfo{year}{2009}) \bibinfo{pages}{2941--2950}.
\bibitem[{Jia and Zhang(2016)}]{jia-2016}
\bibinfo{author}{H.~Jia}, \bibinfo{author}{P.~Zhang}, \bibinfo{journal}{Chem. Eng. J.} \bibinfo{volume}{285} (\bibinfo{year}{2016}) \bibinfo{pages}{252--263}.
\bibitem[{Sontti and Atta(2017)}]{sontti2017cfd}
\bibinfo{author}{S.~G. Sontti}, \bibinfo{author}{A.~Atta}, \bibinfo{journal}{Ind. Eng. Chem. Res.} \bibinfo{volume}{56} (\bibinfo{year}{2017}) \bibinfo{pages}{7401--7412}.
\bibitem[{Nghe et~al.(2011)Nghe, Terriac, Schneider, Li, Cloitre, Abecassis, and Tabeling}]{nghe2011microfluidics}
\bibinfo{author}{P.~Nghe}, \bibinfo{author}{E.~Terriac}, \bibinfo{author}{M.~Schneider}, \bibinfo{author}{Z.~Li}, \bibinfo{author}{M.~Cloitre}, \bibinfo{author}{B.~Abecassis}, \bibinfo{author}{P.~Tabeling}, \bibinfo{journal}{Lab. Chip} \bibinfo{volume}{11} (\bibinfo{year}{2011}) \bibinfo{pages}{788--794}.
\bibitem[{Shao et~al.(2010)Shao, Gavriilidis, and Angeli}]{shao2010mass}
\bibinfo{author}{N.~Shao}, \bibinfo{author}{A.~Gavriilidis}, \bibinfo{author}{P.~Angeli}, \bibinfo{journal}{Chem. Eng. J.} \bibinfo{volume}{160} (\bibinfo{year}{2010}) \bibinfo{pages}{873--881}.
\bibitem[{Picchi et~al.(2015)Picchi, Manerba, Correra, Margarone, and Poesio}]{picchi-2015}
\bibinfo{author}{D.~Picchi}, \bibinfo{author}{Y.~Manerba}, \bibinfo{author}{S.~Correra}, \bibinfo{author}{M.~Margarone}, \bibinfo{author}{P.~Poesio}, \bibinfo{journal}{Int. J. Multiphase Flow} \bibinfo{volume}{73} (\bibinfo{year}{2015}) \bibinfo{pages}{217--226}.
\bibitem[{Laborie et~al.(2015)Laborie, Rouyer, Angelescu, and Lorenceau}]{labor-2015}
\bibinfo{author}{B.~Laborie}, \bibinfo{author}{F.~Rouyer}, \bibinfo{author}{D.~E. Angelescu}, \bibinfo{author}{E.~Lorenceau}, \bibinfo{journal}{Phys. Rev. Lett.} \bibinfo{volume}{114} (\bibinfo{year}{2015}) \bibinfo{pages}{204501}.
\bibitem[{Mansour et~al.(2015)Mansour, Kawahara, and Sadatomi}]{mansour2015}
\bibinfo{author}{M.~H. Mansour}, \bibinfo{author}{A.~Kawahara}, \bibinfo{author}{M.~Sadatomi}, \bibinfo{journal}{Int. J. Multiphase Flow} \bibinfo{volume}{72} (\bibinfo{year}{2015}) \bibinfo{pages}{263--274}.
\bibitem[{Li et~al.(2002)Li, Mouline, and Midoux}]{li2002}
\bibinfo{author}{H.~Z. Li}, \bibinfo{author}{Y.~Mouline}, \bibinfo{author}{N.~Midoux}, \bibinfo{journal}{Chem. Eng. Sci.} \bibinfo{volume}{57} (\bibinfo{year}{2002}) \bibinfo{pages}{339--346}.
\bibitem[{Tang et~al.(2012)Tang, Lu, Zhang, Wang, and Tao}]{tang2012}
\bibinfo{author}{G.~Tang}, \bibinfo{author}{Y.~Lu}, \bibinfo{author}{S.~Zhang}, \bibinfo{author}{F.~Wang}, \bibinfo{author}{W.~Tao}, \bibinfo{journal}{J. Non-Newtonian Fluid Mech.} \bibinfo{volume}{173} (\bibinfo{year}{2012}) \bibinfo{pages}{21--29}.
\bibitem[{Fu et~al.(2011)Fu, Ma, Funfschilling, and Li}]{fu-2011}
\bibinfo{author}{T.~Fu}, \bibinfo{author}{Y.~Ma}, \bibinfo{author}{D.~Funfschilling}, \bibinfo{author}{H.~Z. Li}, \bibinfo{journal}{Chem. Eng. Process. Process Intensif.} \bibinfo{volume}{50} (\bibinfo{year}{2011}) \bibinfo{pages}{438--442}.
\bibitem[{Wang et~al.(2011)Wang, Huang, He, and Chen}]{wanga-2011}
\bibinfo{author}{S.~Wang}, \bibinfo{author}{J.~Huang}, \bibinfo{author}{K.~He}, \bibinfo{author}{J.~Chen}, \bibinfo{journal}{Int. J. Multiphase Flow} \bibinfo{volume}{37} (\bibinfo{year}{2011}) \bibinfo{pages}{1129--1134}.
\bibitem[{Fu et~al.(2012)Fu, Ma, Funfschilling, Zhu, and Li}]{fu-2012b}
\bibinfo{author}{T.~Fu}, \bibinfo{author}{Y.~Ma}, \bibinfo{author}{D.~Funfschilling}, \bibinfo{author}{C.~Zhu}, \bibinfo{author}{H.~Z. Li}, \bibinfo{journal}{AIChE J.} \bibinfo{volume}{58} (\bibinfo{year}{2012}) \bibinfo{pages}{3560--3567}.
\bibitem[{Laborie et~al.(2016)Laborie, Rouyer, Angelescu, and Lorenceau}]{laborie2016}
\bibinfo{author}{B.~Laborie}, \bibinfo{author}{F.~Rouyer}, \bibinfo{author}{D.~E. Angelescu}, \bibinfo{author}{E.~Lorenceau}, \bibinfo{journal}{Soft Matter} \bibinfo{volume}{12} (\bibinfo{year}{2016}) \bibinfo{pages}{9355--9363}.
\bibitem[{Chen et~al.(2013)Chen, Guo, Li, Wang et~al.}]{chen2013}
\bibinfo{author}{B.~Chen}, \bibinfo{author}{F.~Guo}, \bibinfo{author}{G.~Li}, \bibinfo{author}{P.~Wang}, et~al., \bibinfo{journal}{Chem. Eng. Technol.} \bibinfo{volume}{36} (\bibinfo{year}{2013}) \bibinfo{pages}{2087--2100}.
\bibitem[{Harvie et~al.(2006)Harvie, Davidson, and Rudman}]{harvie2006}
\bibinfo{author}{D.~J. Harvie}, \bibinfo{author}{M.~Davidson}, \bibinfo{author}{M.~Rudman}, \bibinfo{journal}{Appl. Math. Modell.} \bibinfo{volume}{30} (\bibinfo{year}{2006}) \bibinfo{pages}{1056--1066}.
\bibitem[{Popinet and Zaleski(1999)}]{popinet1999}
\bibinfo{author}{S.~Popinet}, \bibinfo{author}{S.~Zaleski}, \bibinfo{journal}{Int. J. Numer. Meth. Fluids} \bibinfo{volume}{30} (\bibinfo{year}{1999}) \bibinfo{pages}{775--793}.
\bibitem[{Aulisa et~al.(2006)Aulisa, Manservisi, and Scardovelli}]{aulisa2006}
\bibinfo{author}{E.~Aulisa}, \bibinfo{author}{S.~Manservisi}, \bibinfo{author}{R.~Scardovelli}, \bibinfo{journal}{Comput. Methods in Appl. Mech. Eng.} \bibinfo{volume}{195} (\bibinfo{year}{2006}) \bibinfo{pages}{6239--6257}.
\bibitem[{Francois et~al.(2006)Francois, Cummins, Dendy, Kothe, Sicilian, and Williams}]{francois2006}
\bibinfo{author}{M.~M. Francois}, \bibinfo{author}{S.~J. Cummins}, \bibinfo{author}{E.~D. Dendy}, \bibinfo{author}{D.~B. Kothe}, \bibinfo{author}{J.~M. Sicilian}, \bibinfo{author}{M.~W. Williams}, \bibinfo{journal}{J. Comput. Phys.} \bibinfo{volume}{213} (\bibinfo{year}{2006}) \bibinfo{pages}{141--173}.
\bibitem[{Popinet(2009)}]{popinet2009}
\bibinfo{author}{S.~Popinet}, \bibinfo{journal}{J. Comput. Phys.} \bibinfo{volume}{228} (\bibinfo{year}{2009}) \bibinfo{pages}{5838--5866}.
\bibitem[{Guo et~al.(2015)Guo, Fletcher, and Haynes}]{guo2015}
\bibinfo{author}{Z.~Guo}, \bibinfo{author}{D.~F. Fletcher}, \bibinfo{author}{B.~S. Haynes}, \bibinfo{journal}{Appl. Math. Modell.} \bibinfo{volume}{39} (\bibinfo{year}{2015}) \bibinfo{pages}{4665--4686}.
\bibitem[{Sussman and Puckett(2000)}]{sussman2000}
\bibinfo{author}{M.~Sussman}, \bibinfo{author}{E.~G. Puckett}, \bibinfo{journal}{J. Comput. Phys.} \bibinfo{volume}{162} (\bibinfo{year}{2000}) \bibinfo{pages}{301--337}.
\bibitem[{Keshavarzi et~al.(2014)Keshavarzi, Pawell, Barber, and Yeoh}]{keshavarzi2014}
\bibinfo{author}{G.~Keshavarzi}, \bibinfo{author}{R.~S. Pawell}, \bibinfo{author}{T.~J. Barber}, \bibinfo{author}{G.~H. Yeoh}, \bibinfo{journal}{Chem. Eng. Sci.} \bibinfo{volume}{112} (\bibinfo{year}{2014}) \bibinfo{pages}{25--34}.
\bibitem[{Buwa et~al.(2007)Buwa, Gerlach, Durst, and Schl{\"u}cker}]{buwa2007}
\bibinfo{author}{V.~V. Buwa}, \bibinfo{author}{D.~Gerlach}, \bibinfo{author}{F.~Durst}, \bibinfo{author}{E.~Schl{\"u}cker}, \bibinfo{journal}{Chem. Eng. Sci.} \bibinfo{volume}{62} (\bibinfo{year}{2007}) \bibinfo{pages}{7119--7132}.
\bibitem[{Ray et~al.(2015)Ray, Biswas, and Sharma}]{ray2015}
\bibinfo{author}{B.~Ray}, \bibinfo{author}{G.~Biswas}, \bibinfo{author}{A.~Sharma}, \bibinfo{journal}{J. Fluid Mech.} \bibinfo{volume}{768} (\bibinfo{year}{2015}) \bibinfo{pages}{492--523}.
\bibitem[{Dang et~al.(2015)Dang, Yue, and Chen}]{dang-2015}
\bibinfo{author}{M.~Dang}, \bibinfo{author}{J.~Yue}, \bibinfo{author}{G.~Chen}, \bibinfo{journal}{Chem. Eng. J.} \bibinfo{volume}{262} (\bibinfo{year}{2015}) \bibinfo{pages}{616--627}.
\bibitem[{Chakraborty et~al.(2016)Chakraborty, Rubio-Rubio, Sevilla, and Gordillo}]{chakraborty2016}
\bibinfo{author}{I.~Chakraborty}, \bibinfo{author}{M.~Rubio-Rubio}, \bibinfo{author}{A.~Sevilla}, \bibinfo{author}{J.~Gordillo}, \bibinfo{journal}{Int. J. Multiphase Flow} \bibinfo{volume}{84} (\bibinfo{year}{2016}) \bibinfo{pages}{54--65}.
\bibitem[{Mino et~al.(2016)Mino, Kagawa, Ishigami, and Matsuyama}]{mino2016}
\bibinfo{author}{Y.~Mino}, \bibinfo{author}{Y.~Kagawa}, \bibinfo{author}{T.~Ishigami}, \bibinfo{author}{H.~Matsuyama}, \bibinfo{journal}{Colloids Surf., A} \bibinfo{volume}{491} (\bibinfo{year}{2016}) \bibinfo{pages}{70--77}.
\bibitem[{Kagawa et~al.(2014)Kagawa, Ishigami, Hayashi, Fuse, Mino, and Matsuyama}]{kagawa2014}
\bibinfo{author}{Y.~Kagawa}, \bibinfo{author}{T.~Ishigami}, \bibinfo{author}{K.~Hayashi}, \bibinfo{author}{H.~Fuse}, \bibinfo{author}{Y.~Mino}, \bibinfo{author}{H.~Matsuyama}, \bibinfo{journal}{Soft Matter} \bibinfo{volume}{10} (\bibinfo{year}{2014}) \bibinfo{pages}{7985--7992}.
\bibitem[{Fan et~al.(2014)Fan, Sun, and Chen}]{fan2014}
\bibinfo{author}{W.~Fan}, \bibinfo{author}{Y.~Sun}, \bibinfo{author}{H.~Chen}, \bibinfo{journal}{Chem. Eng. Technol.} \bibinfo{volume}{37} (\bibinfo{year}{2014}) \bibinfo{pages}{1566--1574}.
\bibitem[{Fan et~al.(2016)Fan, Qi, Sun, Zhu, and Chen}]{fan2016}
\bibinfo{author}{W.~Fan}, \bibinfo{author}{T.~Qi}, \bibinfo{author}{Y.~Sun}, \bibinfo{author}{P.~Zhu}, \bibinfo{author}{H.~Chen}, \bibinfo{journal}{Chem. Eng. Technol.} \bibinfo{volume}{39} (\bibinfo{year}{2016}) \bibinfo{pages}{1895--1902}.
\bibitem[{Sussman et~al.(1994)Sussman, Smereka, and Osher}]{suss-1994}
\bibinfo{author}{M.~Sussman}, \bibinfo{author}{P.~Smereka}, \bibinfo{author}{S.~Osher}, \bibinfo{journal}{J. Comput. Phys.} \bibinfo{volume}{114} (\bibinfo{year}{1994}) \bibinfo{pages}{146--159}.
\bibitem[{Hirt and Nichols(1981)}]{hirt-1981}
\bibinfo{author}{C.~W. Hirt}, \bibinfo{author}{B.~D. Nichols}, \bibinfo{journal}{J. Comput. Phys.} \bibinfo{volume}{39} (\bibinfo{year}{1981}) \bibinfo{pages}{201--225}.
\bibitem[{Brackbill et~al.(1992)Brackbill, Kothe, and Zemach}]{brack-1992}
\bibinfo{author}{J.~Brackbill}, \bibinfo{author}{D.~B. Kothe}, \bibinfo{author}{C.~Zemach}, \bibinfo{journal}{J. Comput. Phys.} \bibinfo{volume}{100} (\bibinfo{year}{1992}) \bibinfo{pages}{335--354}.
\bibitem[{Fluent(2017)}]{fluent}
\bibinfo{author}{A.~Fluent}, \bibinfo{journal}{ANSYS FLUENT Inc.}  (\bibinfo{year}{2017}).
\bibitem[{Sussman et~al.(1999)Sussman, Almgren, Bell, Colella, Howell, and Welcome}]{sussman1999}
\bibinfo{author}{M.~Sussman}, \bibinfo{author}{A.~S. Almgren}, \bibinfo{author}{J.~B. Bell}, \bibinfo{author}{P.~Colella}, \bibinfo{author}{L.~H. Howell}, \bibinfo{author}{M.~L. Welcome}, \bibinfo{journal}{J. Comput. Phys.} \bibinfo{volume}{148} (\bibinfo{year}{1999}) \bibinfo{pages}{81--124}.
\bibitem[{Chhabra and Richardson(2011)}]{chhabra2011}
\bibinfo{author}{R.~P. Chhabra}, \bibinfo{author}{J.~F. Richardson}, \bibinfo{title}{Non-Newtonian flow and applied rheology: engineering applications}, \bibinfo{publisher}{Butterworth-Heinemann}, \bibinfo{year}{2011}.
\bibitem[{Issa(1986)}]{issa1986}
\bibinfo{author}{R.~I. Issa}, \bibinfo{journal}{J. Comput. Phys.} \bibinfo{volume}{62} (\bibinfo{year}{1986}) \bibinfo{pages}{40--65}.
\bibitem[{Barth and Jespersen(1989)}]{barth1989}
\bibinfo{author}{T.~Barth}, \bibinfo{author}{D.~Jespersen}, in: \bibinfo{booktitle}{27th Aerospace sciences meeting}, p. \bibinfo{pages}{366}. \DOIprefix\doi{10.2514/6.1989-366}.
\bibitem[{Holt(2012)}]{holt2012}
\bibinfo{author}{M.~Holt}, \bibinfo{title}{Numerical methods in fluid dynamics}, \bibinfo{publisher}{Springer Science \& Business Media}, \bibinfo{year}{2012}.
\bibitem[{Fu et~al.(2016)Fu, Carrier, Funfschilling, Ma, and Li}]{fu2016newtonian}
\bibinfo{author}{T.~Fu}, \bibinfo{author}{O.~Carrier}, \bibinfo{author}{D.~Funfschilling}, \bibinfo{author}{Y.~Ma}, \bibinfo{author}{H.~Z. Li}, \bibinfo{journal}{Chem. Eng. Technol.} \bibinfo{volume}{39} (\bibinfo{year}{2016}) \bibinfo{pages}{987--992}.
\bibitem[{Vayssade et~al.(2014)Vayssade, Lee, Terriac, Monti, Cloitre, and Tabeling}]{vayssade2014}
\bibinfo{author}{A.-L. Vayssade}, \bibinfo{author}{C.~Lee}, \bibinfo{author}{E.~Terriac}, \bibinfo{author}{F.~Monti}, \bibinfo{author}{M.~Cloitre}, \bibinfo{author}{P.~Tabeling}, \bibinfo{journal}{Phys. Rev. E} \bibinfo{volume}{89} (\bibinfo{year}{2014}) \bibinfo{pages}{052309}.
\bibitem[{Madadelahi and Shamloo(2018)}]{madadelahi2018}
\bibinfo{author}{M.~Madadelahi}, \bibinfo{author}{A.~Shamloo}, \bibinfo{journal}{J. Non-Newtonian Fluid Mech.} \bibinfo{volume}{251} (\bibinfo{year}{2018}) \bibinfo{pages}{88--96}.
\bibitem[{Ren et~al.(2015)Ren, Liu, and Shum}]{ren2015}
\bibinfo{author}{Y.~Ren}, \bibinfo{author}{Z.~Liu}, \bibinfo{author}{H.~C. Shum}, \bibinfo{journal}{Lab. Chip} \bibinfo{volume}{15} (\bibinfo{year}{2015}) \bibinfo{pages}{121--134}.
\bibitem[{Cho et~al.(2012)Cho, Chen et~al.}]{cho2012}
\bibinfo{author}{C.-C. Cho}, \bibinfo{author}{C.-L. Chen}, et~al., \bibinfo{journal}{Chem. Eng. J.} \bibinfo{volume}{191} (\bibinfo{year}{2012}) \bibinfo{pages}{132--140}.
\bibitem[{Sang et~al.(2009)Sang, Hong, and Wang}]{sang2009}
\bibinfo{author}{L.~Sang}, \bibinfo{author}{Y.~Hong}, \bibinfo{author}{F.~Wang}, \bibinfo{journal}{Microfluid. Nanofluid.} \bibinfo{volume}{6} (\bibinfo{year}{2009}) \bibinfo{pages}{621--635}.
\bibitem[{Deng et~al.(2017)Deng, Wang, Huang, and Cheng}]{deng2017}
\bibinfo{author}{C.~Deng}, \bibinfo{author}{H.~Wang}, \bibinfo{author}{W.~Huang}, \bibinfo{author}{S.~Cheng}, \bibinfo{journal}{Colloids Surf., A} \bibinfo{volume}{533} (\bibinfo{year}{2017}) \bibinfo{pages}{1--8}.
\bibitem[{Dietrich et~al.(2008)Dietrich, Poncin, Pheulpin, and Li}]{dietrich2008passage}
\bibinfo{author}{N.~Dietrich}, \bibinfo{author}{S.~Poncin}, \bibinfo{author}{S.~Pheulpin}, \bibinfo{author}{H.~Z. Li}, \bibinfo{journal}{AIChE J.} \bibinfo{volume}{54} (\bibinfo{year}{2008}) \bibinfo{pages}{594--600}.
\bibitem[{Fu et~al.(2016)Fu, Ma, and Li}]{fu2016breakup}
\bibinfo{author}{T.~Fu}, \bibinfo{author}{Y.~Ma}, \bibinfo{author}{H.~Z. Li}, \bibinfo{journal}{Chem. Eng. Sci.} \bibinfo{volume}{144} (\bibinfo{year}{2016}) \bibinfo{pages}{75--86}.
\bibitem[{Fu et~al.(2011)Fu, Ma, Funfschilling, and Li}]{fu2011gas}
\bibinfo{author}{T.~Fu}, \bibinfo{author}{Y.~Ma}, \bibinfo{author}{D.~Funfschilling}, \bibinfo{author}{H.~Z. Li}, \bibinfo{journal}{Microfluid. Nanofluid.} \bibinfo{volume}{10} (\bibinfo{year}{2011}) \bibinfo{pages}{1135--1140}.
\bibitem[{Bretherton(1961)}]{breth-1961}
\bibinfo{author}{F.~Bretherton}, \bibinfo{journal}{J. Fluid Mech.} \bibinfo{volume}{10} (\bibinfo{year}{1961}) \bibinfo{pages}{166--188}.
\bibitem[{Bai et~al.(2016)Bai, Fu, Zhao, and Cheng}]{bai2016dro}
\bibinfo{author}{L.~Bai}, \bibinfo{author}{Y.~Fu}, \bibinfo{author}{S.~Zhao}, \bibinfo{author}{Y.~Cheng}, \bibinfo{journal}{Chem. Eng. Sci.} \bibinfo{volume}{145} (\bibinfo{year}{2016}) \bibinfo{pages}{141--148}.
\bibitem[{Roumpea et~al.(2017)Roumpea, Chinaud, and Angeli}]{roumpea2017experimental}
\bibinfo{author}{E.~Roumpea}, \bibinfo{author}{M.~Chinaud}, \bibinfo{author}{P.~Angeli}, \bibinfo{journal}{AIChE J.} \bibinfo{volume}{63} (\bibinfo{year}{2017}) \bibinfo{pages}{3599--3609}.
\bibitem[{Lan et~al.(2015)Lan, Li, and Luo}]{lan2015numerical}
\bibinfo{author}{W.~Lan}, \bibinfo{author}{S.~Li}, \bibinfo{author}{G.~Luo}, \bibinfo{journal}{Chem. Eng. Sci.} \bibinfo{volume}{134} (\bibinfo{year}{2015}) \bibinfo{pages}{76--85}.
\bibitem[{Hernainz and Caro(2002)}]{hernainz2002}
\bibinfo{author}{F.~Hernainz}, \bibinfo{author}{A.~Caro}, \bibinfo{journal}{Colloids Surf., A} \bibinfo{volume}{196} (\bibinfo{year}{2002}) \bibinfo{pages}{19--24}.
\bibitem[{Xu et~al.(2006)Xu, Li, Wang, and Luo}]{xu2006}
\bibinfo{author}{J.~Xu}, \bibinfo{author}{S.~Li}, \bibinfo{author}{Y.~Wang}, \bibinfo{author}{G.~Luo}, \bibinfo{journal}{Appl. Phys. Lett.} \bibinfo{volume}{88} (\bibinfo{year}{2006}) \bibinfo{pages}{133506}.
\bibitem[{Dietrich et~al.(2008)Dietrich, Poncin, Midoux, and Li}]{dietri-2008}
\bibinfo{author}{N.~Dietrich}, \bibinfo{author}{S.~Poncin}, \bibinfo{author}{N.~Midoux}, \bibinfo{author}{H.~Z. Li}, \bibinfo{journal}{Langmuir} \bibinfo{volume}{24} (\bibinfo{year}{2008}) \bibinfo{pages}{13904--13911}.

\end{thebibliography}

\end{document}